\documentclass[nofootinbib,aps,prl,preprint]{revtex4-1}
\usepackage{amsmath}
\usepackage{bbold}
\usepackage{graphicx}
\usepackage[hidelinks]{hyperref}
\usepackage{bookmark}
\usepackage{lineno}
\newcommand{\etal}{\textit{et al}.}

\newcommand{\parallelsum}{\mathbin{\!/\mkern-5mu/\!}}
\newcommand{\beginsupplement}{%
        \setcounter{table}{0}
        \renewcommand{\thetable}{S\arabic{table}}%
        \setcounter{figure}{0}
        \renewcommand{\thefigure}{S\arabic{figure}}%
     }

\begin{document}
\title{Topological Nernst effect of the two-dimensional skyrmion lattice}
\author{Max Hirschberger$^{1,2}$}\email{hirschberger@ap.t.u-tokyo.ac.jp}
\author{Leonie Spitz$^{1}$}\altaffiliation{Current address: Physik-Department, Technical University of Munich, 85748 Garching, Germany}
\author{Takuya Nomoto$^{2}$}
\author{Takashi Kurumaji$^{1}$}\altaffiliation{Current address: Department of Physics, Massachusetts Institute of Technology, Cambridge, MA 02139, USA}
\author{Shang Gao$^{1}$}\altaffiliation{Current address: Materials Science \& Technology Division and Neutron Science Division, Oak Ridge National Laboratory, Oak Ridge, TN 37831, USA}
\author{Jan Masell$^{1}$}
\author{Taro Nakajima$^{1}$}\altaffiliation{Current address: Institute for Solid State Physics, The University of Tokyo, Kashiwa, Chiba 277-8561, Japan}
\author{Akiko Kikkawa$^{1}$}
\author{Yuichi Yamasaki$^{3,4}$}
\author{Hajime Sagayama$^{5}$}
\author{Hironori Nakao$^{5}$}
\author{Yasujiro Taguchi$^{1}$}
\author{Ryotaro Arita$^{2}$}
\author{Taka-hisa Arima$^{1,6}$}
\author{Yoshinori Tokura$^{1,2,7}$}
\date{\today}
\affiliation{$^{1}$RIKEN Center for Emergent Matter Science (CEMS), Wako, Saitama 351-0198, Japan}
\affiliation{$^{2}$Department of Applied Physics and Quantum-Phase Electronics Center, The University of Tokyo, Bunkyo-ku, Tokyo 113-8656, Japan}
\affiliation{$^{3}$Research and Services Division of Materials Data and Integrated System (MaDIS), National Institute for Materials Science (NIMS), Tsukuba, Ibaraki 305-0047, Japan}
\affiliation{$^{4}$PRESTO, Japan Science and Technology Agency (JST), Kawaguchi, Saitama 332-0012, Japan}
\affiliation{$^{5}$Institute of Materials Structure Science, High Energy Accelerator Research Organization, Tsukuba, Ibaraki 305-0801, Japan}
\affiliation{$^{6}$Department of Advanced Materials Science, The University of Tokyo, Kashiwa, Chiba 277-8561, Japan}
\affiliation{$^{7}$Tokyo College, The University of Tokyo, Bunkyo-ku, Tokyo 113-8656, Japan}

\begin{abstract}
The topological Hall effect (THE) and its thermoelectric counterpart, the topological Nernst effect (TNE), are hallmarks of the skyrmion lattice phase (SkL). We observed the giant TNE of the SkL in centrosymmetric Gd$_2$PdSi$_3$, comparable in magnitude to the largest anomalous Nernst signals in ferromagnets. Significant enhancement (suppression) of the THE occurs when doping electrons (holes) to Gd$_2$PdSi$_3$. On the electron-doped side, the topological Hall conductivity approaches the characteristic threshold $\sim 1000\,\left(\mathrm{\Omega cm}\right)^{-1}$ for the intrinsic regime. We use the filling-controlled samples to confirm Mott's relation between TNE and THE and discuss the importance of Gd-5d orbitals for transport in this compound.
\end{abstract}

\maketitle
A skyrmion spin-vortex \cite{Bogdanov1989,Muehlbauer2009} represents a quantized unit of scalar spin chirality $\chi_{\alpha\beta\gamma}=\textbf{S}_\alpha\cdot\left(\mathbf{S}_\beta \times \mathbf{S}_\gamma \right)$, defined for three neighboring magnetic moments on lattice sites $\alpha$, $\beta$, and $\gamma$. It was realized early on that spin-winding results in an emergent gauge field acting on moving particles, leading to anomalies such as the topological (or geometrical) Hall effect (THE) and its sibling, the topological Nernst effect (TNE) \cite{Ye1999,Ohgushi2000,Shindou2001,Bruno2004,Shiomi2013,Mizuta2016,Mizuta2018}. The relative magnitude of the carrier mean-free path $l_\text{mfp}$ as compared to the size of the skyrmion $\lambda_\text{sk}$, i.e. to the size of the magnetic unit cell, governs the appropriate starting point for theoretical modeling\cite{Onoda2004}: Well-known cases of non-centrosymmetric materials with skyrmion lattice (SkL) phase, such as MnSi\cite{Muehlbauer2009,Lee2009,Neubauer2009,Ritz2013}, fall into the regime $l_\text{mfp} \ll \lambda_\text{sk}$ and it is understood that the (weak) THE relates to a Berry-phase induced deflection of wavepackets moving through the twisted spin texture in real space \cite{Kawamura2003,Bruno2004,Ritz2013,Nakazawa2018}. Meanwhile, the as-yet unexplored 'intrinsic' (momentum-space) limit $l_\text{mfp} \ge \lambda_\text{sk}$ necessitates a modification of the electronic wave functions themselves by the presence of magnetic order, predicted to yield a large THE and TNE due to Berry curvature in reciprocal space\cite{Shindou2001,Martin2008,Hamamoto2015,Lado2015,Goebel2018,Wang2019}.

To describe the connection between TNE and THE, we write the electric currents $\mathbf{J}$ emanating from an applied electric field $\mathbf{E}$ or an applied temperature gradient $\left(-\nabla T\right)$ as $J_i=\sigma_{ij} E_j$ and $J_i=\alpha_{ij} \left(-\nabla_j T\right)$, respectively. The TNE provides insight into the effect of a variation of the chemical potential $\zeta$, as expressed by the Mott relation \cite{Ziman1979,Smrcka1977} 
\begin{equation}
\label{eq:mott}
\alpha_{ij}/T=-\left(\pi^2/3\right) \left(k_B^2/e\right) \left(\partial \sigma_{ij}/\partial \varepsilon\right)_{\varepsilon=\zeta}
\end{equation}
where $k_B$, $e$ ($>0$), and $\varepsilon$ represent the Boltzmann constant, the fundamental charge, and the band filling energy, respectively. Due to experimental constraints such as relatively weak spin polarization and low skyrmion density in the ambient pressure, equilibrium SkL phases of chiral B20 compounds (skyrmion-skyrmion distance $\sim 20-200$ nanometers), the TNE of a SkL has never hitherto been observed experimentally. 

In this letter, we report the giant TNE of the SkL in centrosymmetric Gd$_2$PdSi$_3$, compare its magnitude to recently observed giant anomalous Nernst responses in ferromagnets, and show how the TNE is related to the enhancement (decrease) of the THE under electron (hole) doping through the Mott relation. We demonstrate that on the electron-doped side, the intrinsic limit of the THE may be within reach, and that the large transport responses in Gd$_2$PdSi$_3$ are likely related to the prevalence of Gd-5d conducting orbitals in close proximity to the Fermi energy.

\begin{figure*}[!htb]
  \begin{center}
		\includegraphics[width=1.0\linewidth]{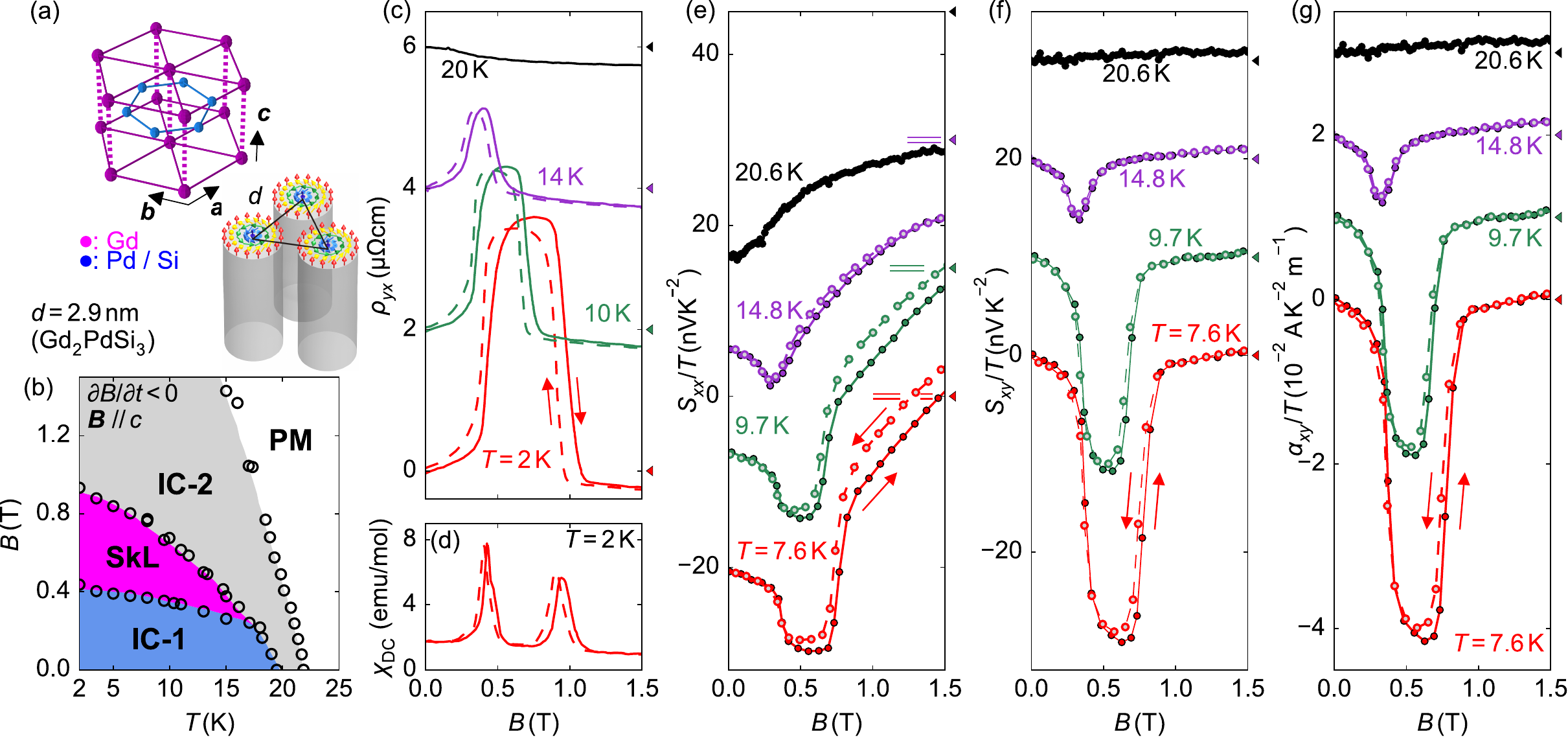}
    \caption[]{(color online). (a) Hexagonal structure (basic AlB$_2$-type) of Gd$_2$PdSi$_3$. Inset: high-density SkL state with nanometer-sized inter-skyrmion distance $d$. (b) Magnetic phase diagram for $\mathbf{B}\,\parallelsum\, c$ and decreasing field $\partial B/\partial t<0$, as adapted from Ref. \cite{Hirschberger2020}. Labels indicate the IC-1 ground state, the skyrmion lattice phase (SkL), the IC-2 fan-like state, and the paramagnetic regime (PM). (c) Hall resistivity, (e) thermopower, (f) Nernst effect, and (g) Nernst conductivity. For the latter three, entropy factor $\sim T$ was removed. (d) Magnetic susceptibility $\chi_{DC} (B)$ at the lowest $T=2\,$K. Curves in panels (c, e-g) were shifted by vertical offsets of $2\,\mu\Omega$cm, $15\,$nV K$^{-2}$, $10\,$nV K$^{-2}$, and $0.01\,$AK$^{-2}$m$^{-1}$, respectively. The zero-level for each curve is indicated by a colored triangle at the right side of the panel. Solid arrows mark the direction of the field ramp. In (e), red,  green, and purple twinned lines also indicate zero-levels for $T=7.6$, $9.7$, and $14.8\,$K.}
    \label{fig:fig1}
  \end{center}
\end{figure*}

Rare-earth intermetallics with SkL phase, such as Gd$_2$PdSi$_3$\cite{Kurumaji2019}, Gd$_3$Ru$_4$Al$_{12}$\cite{Hirschberger2019}, and GdRu$_2$Si$_2$\cite{Khanh2020} are highly suitable for investigating the transport response from the emergent gauge field: A tiny vortex-vortex distance $\lambda\sim2-3\,$nm leads to giant responses, strongly modifying the trajectory of moving charge carriers. We focus here on centrosymmetric, hexagonal Gd$_2$PdSi$_3$ [Fig. \ref{fig:fig1}(a)], where magnetic long-range order onsets at $T_N\sim 20\,$K. The dominant magnetic ion is Gd$^{3+}$ in the triangular lattice plane. Dzyaloshinskii-Moriya interactions, which are intrinsic to non-centrosymmetric material platforms and which favor helical order and skyrmion spin textures, are expected to be globally absent in this centrosymmetric bulk crystal\cite{Kurumaji2019}. Instead, skyrmion formation is driven by frustrated interactions mediated through the conduction electrons \cite{Inosov2009,Nomoto2020} and remarkably, skyrmions were found to exist (for magnetic field $\mathbf{B}\parallelsum\,\mathbf{c}$-axis) even at the lowest $T = 2\,$K, i.e. at $T/T_N\sim 0.1$ \cite{Kurumaji2019}. Located between two phases with zero net scalar spin chirality [spiral-like IC-1 (possibly multi-$\mathbf{q}$) and the fan-like IC-2, Fig. 1(b)], the topologically stable SkL is bounded by sharp, first-order phase transitions as observed in the strongly hysteretic field-derivative of the DC magnetization [DC susceptibility $\chi_{DC}$, Fig. 1(d) see also Refs. \cite{Saha1999,Kurumaji2019}]. This phase alone was found to host enormous THE and TNE responses in our high-resolution transport experiments [Fig. 1(c, e-g), see Supplementary Information for technical details\cite{SI}], the Hall signal being in good agreement with previous work \cite{Saha1999,Kurumaji2019}. 

\begin{figure}[b]
  \begin{center}
		\includegraphics[width=0.7\linewidth]{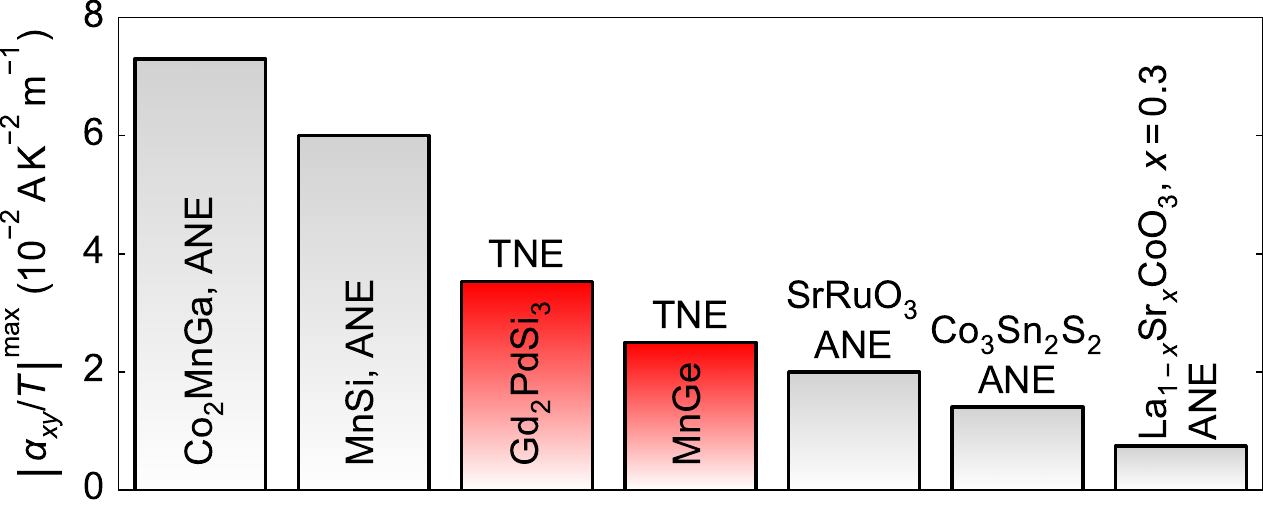}
    \caption[]{(color online). Comparison of large Nernst conductivity $\alpha_{xy}/T$ driven by magnetic order for various materials reported in the literature. Grey shading indicates anomalous Nernst response (ANE, proportional to the net magnetization $M$), while red shading is reserved for the topological Nernst effect (TNE, proportional to the scalar spin chirality).}
    \label{fig:fig2}
  \end{center}
\end{figure}

\begin{figure}[!htb]
  \begin{center}
		\includegraphics[width=0.45\linewidth]{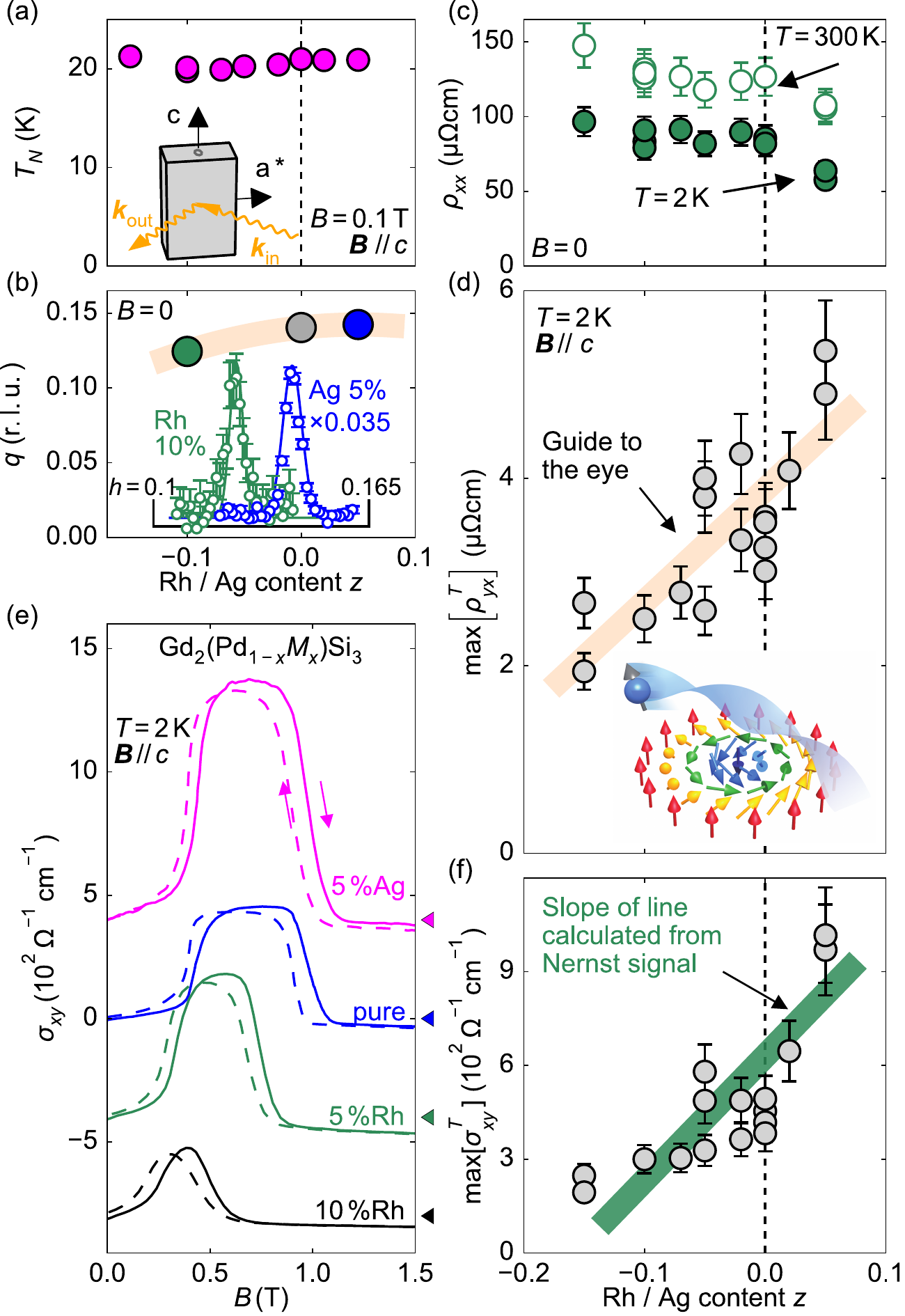}
    \caption[]{(color online). (a) N{\'e}el temperature $T_N$ of Gd$_2$(Pd$_{1-x}M_x$)Si$_3$ from magnetic susceptibility. $M=\,$Rh, Ag corresponds to $z<0$ ($z>0$), respectively (see text). Inset: sample geometry for resonant elastic x-ray diffraction (REXS) at the Gd-L$_2$ edge in reflection geometry. (b) Magnetic wavenumber $q$, determined from REXS. The data point for pure Gd$_2$PdSi$_3$ (grey disk) is reproduced from Ref. \cite{Kurumaji2019}. Inset: Line scans of REXS intensity along $(2-h,1,0)$ in $\mathbf{k}$-space for two samples, after subtraction of a constant background. The beam energy was set to $E = 7.935\,$keV and the sample temperature was $T = 4\,$K ($5\,$K) for the Ag-doped (Rh-doped) crystal. Scale of $y$-axis (not shown) is intensity normalized by monitor counts (arb. units). Statistical errors of detector counts are indicated. (c) Longitudinal resistivity as well as extremal (d) topological Hall resistivity and (f) topological Hall conductivity. Systematic (sample shape) errors are indicated. In (e), raw data of Hall conductivity as a function of magnetic field. Inset of (d), illustration of the real-space Berry-phase mechanism for the THE/TNE\cite{Bruno2004,Neubauer2009}. Red lines in (b, d) are guides to the eye, while the green line in (e) was calculated from the Nernst conductivity (see text). Vertical dashed lines mark $z=0$.}
    \label{fig:fig3}
  \end{center}
\end{figure}
We consider cases where Nernst signals arising from non-coplanar spin arrangements have previously been reported: (i) Pyrochlore Nd$_2$Mo$_2$O$_7$, a canted ferromagnet where the signal is roughly proportional to the net magnetization at all $T$ studied \cite{Hanasaki2008}. (ii) B-20 type, helimagnetic MnGe, where the THE and TNE probe the imbalance between positive and negative contributions to the gauge field. These originate from magnetic monopoles and anti-monopoles, respectively \cite{Kanazawa2011,Shiomi2013}. (iii) Thin films of the Heusler alloy Mn$_{1.8}$PtSn, the magnetic structure of which was not verified independently \cite{Schlitz2019}. In contrast to these cases, the Nernst effect from the SkL laid out here represents a minimal, textbook-like example of the thermoelectric response emerging from a spin texture with integer winding per magnetic unit cell.

The Nernst conductivity (sometimes referred to as transverse thermoelectric conductivity or transverse Peltier conductivity \cite{SI}) is calculated via $\alpha_{xy}=S_{xy} \sigma_{xx}+S_{xx} \sigma_{xy}$, requiring input from the longitudinal thermopower $S_{xx}$ [Fig. 1(e)]. $S_{xx}$ is of comparable magnitude for the two low-field phases IC-1 and SkL; the strong difference of THE and TNE in the respective phases, as well as sharp maxima in $\chi_\text{DC}$ [Fig. \ref{fig:fig1}(d)], clearly distinguish these two states. Figure 1(f,g) shows the experimentally obtained Nernst effect $S_{xy}/T$ and Nernst conductivity $\alpha_{xy}/T$ (entropy factors removed). Note that $\alpha_{xy}/T$ represents the intrinsic transverse thermoelectric response\cite{Ziman1979}, while the derived observable $S_{xy}$ depends strongly on $T$ (via the entropy factor) and the resistivity, i.e. the scattering properties, of the material\cite{Ding2019}. 

For perspective we compare, side-by-side in Fig. 2, representative magnetic materials with anomalous (ANE, i.e. proportional to the net magnetization $M$) or topological (TNE, proportional to the scalar spin chirality) $\alpha_{xy}/T$. Although the TNE of Gd$_2$PdSi$_3$ cannot match the ultra-large ANE exhibited by the ferromagnetic Weyl semimetal Co$_2$MnGa \cite{Sakai2018}, it easily outperforms the ANE of more metallic ferromagnets, e.g. SrRuO$_3$ \cite{Miyasato2007} or elemental Fe and Co, for which only sparse data \cite{Watzmann2016} and calculated values \cite{Weischenberg2013} are available at present \cite{SI}. 

The large TNE signal with negative sign indicates, from Eq. \ref{eq:mott}, that significant enhancement of the THE should be observed when moving the chemical potential $\zeta$ of Gd$_2$PdSi$_3$ upwards. Knowing that Pd-4d orbitals are located about $4\,$eV below $\zeta$\cite{Chaika2001}, a gentle shift of $\zeta$ may be achieved using a series of slightly carrier-doped crystals Gd$_2$(Pd$_{1-x}M_x$)Si$_3$ with $M = \mathrm{Rh}$ (hole-doping) or Ag (electron-doping), as shown in Fig. \ref{fig:fig3}. We define a generalized nominal dopant concentration $z=-x$ on the hole-doped and $z = x$ on the electron-doped side. The single crystals were characterized thoroughly using solid state techniques\cite{SI}. Figure 3(a,b) reports the ordering temperature $T_N$ from magnetic susceptibility as well as the modulation vector $\mathbf{q} = (q,0,0)$ of the ground state from resonant elastic x-ray scattering (REXS) at the Gd-L$_2$ absorption edge in reflection geometry [sketch in Fig. 3(a)]. We determined $q$ from scans of scattering intensity along high-symmetry lines in momentum space [Fig. \ref{fig:fig3}(b), inset]. These data show that the magnetic properties are but weakly affected by chemical substitution on the Pd-site. 

In electrical transport measurements, the topological Hall resistivity $\rho_{yx}^T$ changes significantly with $z$ [Fig. \ref{fig:fig3}(c,d), respectively]. We calculate the topological Hall conductivity $\sigma_{xy}^T = \rho_{yx}^T/\left(\rho_{xx}^2 +\rho_{yx}^2\right)$, shown in Fig. \ref{fig:fig3}(e) for selected samples. The peak value $\text{max}[\sigma_{xy}^T(B$, $T=2\,$K)] is plotted in Fig. \ref{fig:fig3}(f). The green line in this panel has a slope determined directly from the magnitude of the topological Nernst conductivity $\alpha_{xy}^T/T$ using Eq. \ref{eq:mott}\cite{SI}. A necessary ingredient of the calculation, the inverse density of states $\partial \varepsilon/\partial z$, was estimated from the specific heat of iso-electronic Y$_2$PdSi$_3$\cite{Mallik1996,SI}. Good agreement of the doping study with the Nernst signal - without any adjustable parameters - amounts to a direct experimental confirmation of Eq. \ref{eq:mott} for the THE and TNE.

The combined band filling and thermoelectric experiment establishes Gd$_2$PdSi$_3$ as a model material for the quantitative exploration of transport responses in the presence of rather mild changes to the magnetic properties. Note that the Mott relation for the TNE has, to the best of our knowledge, never before been demonstrated in the literature. Even for the ANE, the only test without adjustable parameters was carried out using thin films of Cr$_x$(Sb$_{1-y}$Bi$_y$)$_{2-x}$Te$_3$, with ferromagnetism induced by dilute Cr-spins ($x = 0.15$, Refs. \cite{Guo2017,SI}). We emphasize that not only the sign of the filling-induced change to $\sigma_{xy}^T$ can be predicted from $\alpha_{xy}$, but even its magnitude to within several tens of percents.

\begin{figure}[htb]
  \begin{center}
		\includegraphics[width=0.7\linewidth]{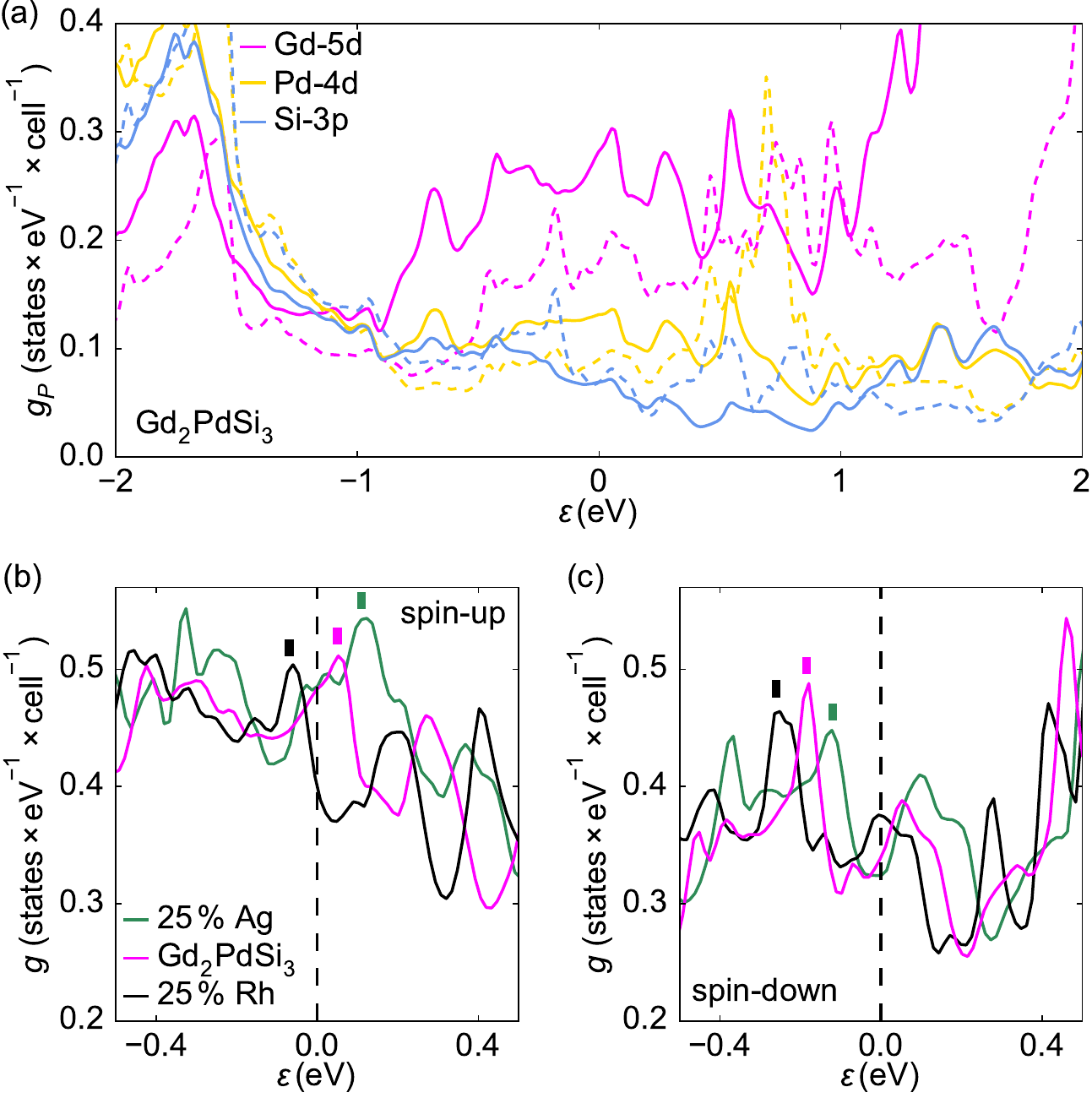}
    \caption[]{(color online). Density functional theory calculations, neglecting Gd-4f, in the spin-polarized state of pure Gd$_2$PdSi$_3$. (a) Partial density of states (P-DOS) $g_P(\varepsilon)$; solid and dashed lines represent spin-up and spin-down bands, respectively. (b,c) Total DOS $g(\varepsilon)$ for Gd$_2$PdSi$_3$ and two doped derivatives. The vertical fat lines are guides to the eye, marking the shift of prominent features in $g(\varepsilon)$ with band filling.}
    \label{fig:fig4}
  \end{center}
\end{figure}

We further scrutinize underlying assumptions made in comparing the TNE with the THE for the doped crystals. Firstly, Eq. \ref{eq:mott} describes the regime of linear response. Hence, we note that despite being moderately large ($-\partial_x T\sim 0.3\,$K/mm), the longitudinal $T$-gradient in our Nernst experiment on Gd$_2$PdSi$_3$ is too small to unpin the SkL, and to result in a flux-flow type Nernst effect\cite{SI}. Secondly, Smr\v{c}ka and Streda derived Eq. \ref{eq:mott} by changing $\zeta$ while leaving all other parameters unaffected - such as character of the magnetic ordering, nature of the electronic bands, and scattering processes\cite{Smrcka1977}. We have demonstrated the stability of the magnetic order in Fig. \ref{fig:fig3} and, because even stoichiometric Gd$_2$PdSi$_3$ already has significant residual resistivity, the introduction of dopants does not excessively affect the scattering properties \cite{SI}. Thirdly, the electronic spectrum of Gd$_2$PdSi$_3$ is sufficiently broadened by disorder to justify the linear approximation [green line in Fig. \ref{fig:fig3}(f)]\cite{SI}. 

It is worth emphasizing that the validity of Mott's relation is in itself experimental evidence for the suitability of the rigid band scenario in describing mild changes of composition in Gd$_2$(Pd$_{1-x}M_x$)Si$_3$ ($M=\,$Rh, Ag). However, the notion of rigid bands warrants some more careful examination. As a first step, we have calculated the partial density of states (P-DOS) $g_P(\varepsilon)$ of pure Gd$_2$PdSi$_3$ in the framework of density functional theory (DFT) [Fig. \ref{fig:fig4}(a)], using the Ce$_2$CoSi$_3$ structure type. The calculations were carried out in the fully spin-polarized state (see Supplementary Information for technical details). Figure \ref{fig:fig4}(b,c) further illustrates the effect of doping. The size of the unit cell for the doped cases was doubled along $c$, and one Pd atom was replaced with Rh or Ag\cite{SI}. Although this is an imperfect approximation for randomly distributed dopants in the experimental study, shifting of the total density of states (DOS) to the left (right) side for electron (hole) doping was observed in these calculations, consistent with the rigid band scheme.

The P-DOS in Fig. \ref{fig:fig4}(a) is consistent with previous photoemission work on $R_2$PdSi$_3$ ($R=\,$Tb, La, Gd)\cite{Chaika2001} and indicates that Gd-5d orbitals dominate the total DOS at $\zeta$. Sizable Hund's rule coupling within the atomic shell of Gadolinium underpins the large THE and TNE in this material. Considering the giant $\sigma_{xy}^T$ on the electron-doped side in Fig. \ref{fig:fig3}, it is instructive to examine key material parameters. We draw on experimental data of the carrier mobility (from the normal Hall effect) and $\rho_{xx}$ to estimate the carrier mean free path in stoichiometric Gd$_2$PdSi$_3$\cite{SI}. Under the assumption of a two-fold spin degenerate spherical (tubular) Fermi surface, $l_\text{mfp} = 3.9\,$nm ($=2.8\,$nm). Despite disorder, this is comparable to or larger than the characteristic dimensions of the spin texture. Moreover, an order-of-magnitude estimate for the typical Hall conductivity in the intrinsic region yields $\sigma_{xy}^\text{int}\approx \,950\,\mathrm{\Omega}^{-1}\mathrm{cm}^{-1}$, not far from the present experimental result\cite{SI}.

In conclusion, the insights presented here not only demonstrate a giant TNE from skyrmions and establish the validity of Mott's relation for THE and TNE, but also provide a guiding post for driving deeper into the intrinsic regime of THE and TNE for centrosymmetric skyrmion hosts. The present Nernst response of high-density skyrmion textures in the centrosymmetric magnet Gd$_2$PdSi$_3$ is on par with the largest anomalous Nernst signals ever observed in ferromagnets.

\textit{Acknowledgments.} We are indebted to H. Ishizuka, M.S. Bahramy, N. Nagaosa, F. Kagawa, and H. Oike for enlightening discussions. L.S. was funded by the German Academic Exchange Service (DAAD) via a PROMOS scholarship awarded by the German Federal Ministry of Education and Research (BMBF), with financial and administrative support by C. Pfleiderer. Work at RIKEN was carried out under JST CREST Grant Number JPMJCR1874 (Japan). M. H. and J. M. were supported as Humboldt/JSPS International Research Fellows (18F18804 and 19F19815, respectively). Contributions by M. Ishida to the graphical presentation of the figures are gratefully acknowledged. This work was performed under the approval of the Photon Factory Program Advisory Committee (Proposal No. 2018G570).

\newpage
\section{Supplementary Material}
\beginsupplement

\section[Characterization of crystal structures by x-ray powder diffraction]{Characterization of crystal structures by x-ray powder diffraction}

\begin{figure*}[htb]
  \centering
  \includegraphics[width=0.85\linewidth]{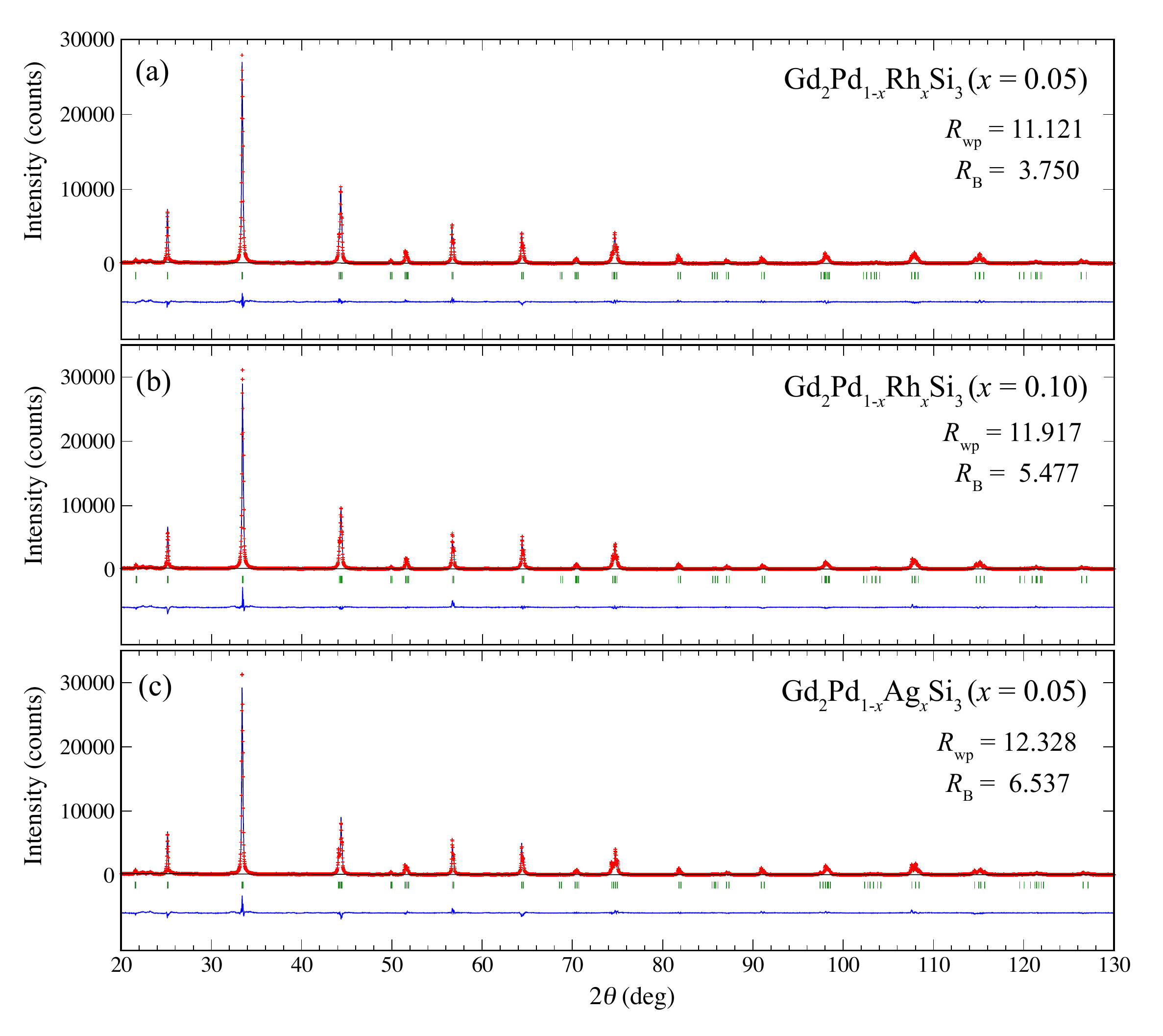}
  \caption{Refinement of x-ray powder patterns for representative samples of slightly doped Gd$_2$PdSi$_3$. Red dots are measured data, blue line is the fit curve, blue crosses are difference between fit and data, green lines indicate reflections expected from a numerical simulation of the powder spectrum. $R_{wp}$ is the $R$-factor for the whole pattern including background, while $R_B$ is the $R$-factor calculated only in the close vicinity of the Bragg peaks. Due to the limited number of peaks in the hexagonal structure, $R_B$ is a better measure for the quality of the fit.}
\label{fig:si_xrd}
\end{figure*}
Single crystals were grown by the optical floating zone technique. Initially, stoichiometric combinations of elemental Gd, Pd and Si (as well as small amounts of Rh and Ag) were melted in an arc furnace. The pellets (mass $3-5$ grams) were turned over and re-melted at least five times from both sides in this step, to ensure homogeneity. We found that it was important to control the temperature in an intermediate range during this step. The surface of the molten pellet serves as visual indicator: strong vibration of the red-hot surface indicates excess heating, which we found to be correlated with evaporation of Si. We further note that the buttons obtained from successful arc-melting typically show a small dimple on the upper side of the pellet.

\begin{figure*}[htb]
  \centering
  \includegraphics[width=1.0\linewidth]{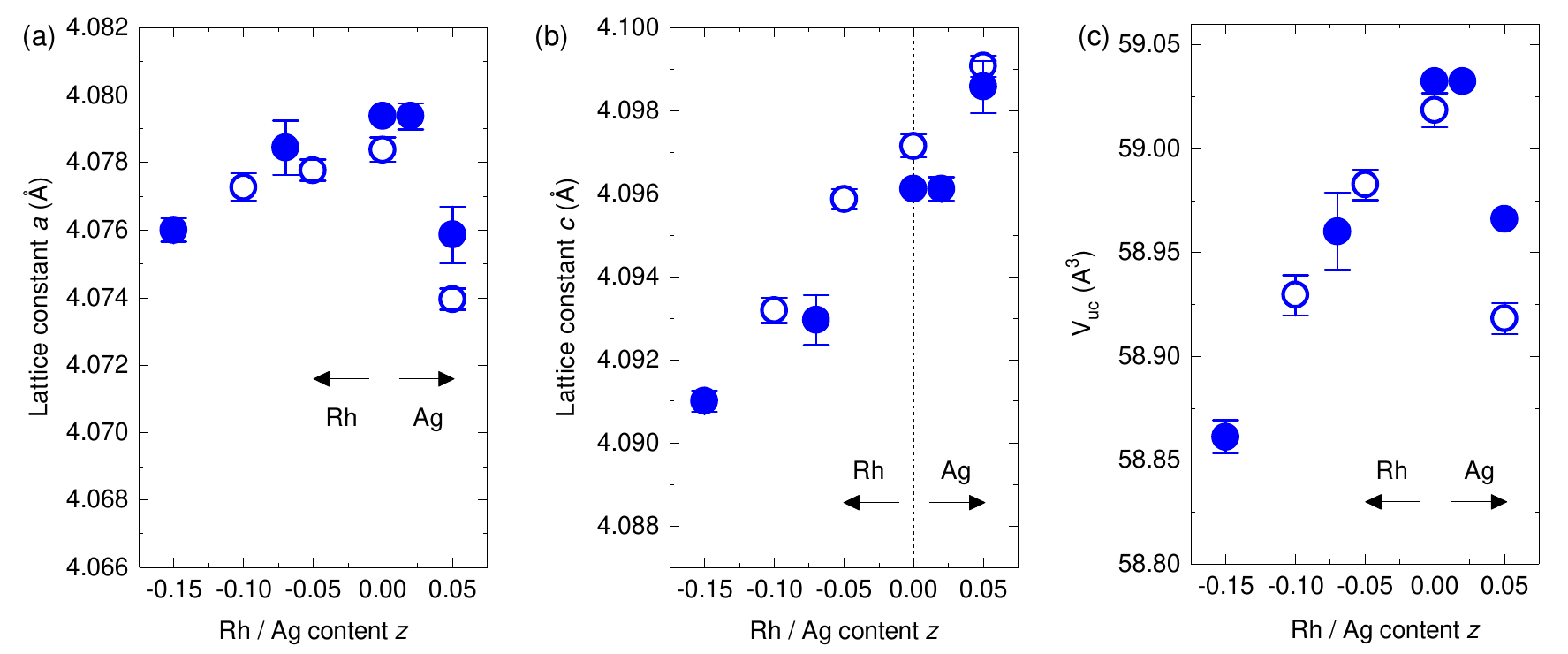}
  \caption{Hexagonal lattice constants $a$, $c$ as well as unit cell volume $V_{uc}$ extracted from x-ray powder patterns for slightly doped Gd$_2$PdSi$_3$. Solid and open disks indicate crystals grown under slightly different conditions (temperature in initial arc-melting step was higher for solid symbols).}
\label{fig:si_lattice_constants}
\end{figure*}

The resulting polycrystals were crushed in a tungsten carbide mortar. Powder-xray diffraction (XRD) was carried out on selected pieces at this stage, to ensure the AlB$_2$-type phase formed in the arc-melting step. Next, a rod-shaped polycrystal was prepared from the crushed pieces in a second arc-melting procedure. The rod was mounted into a halogen-lamp based optical floating zone furnace equipped for high vacuum operation. The sample space was evacuated for $12$ hours before starting the growth, which was carried out under Argon flow at growth speeds of $2-6\,$mm/hour. The single crystals were cut and oriented using Laue back-reflection. Their single-phase nature and stoichiometry was characterized by powder-XRD of crushed pieces, as well as energy-dispersive x-ray spectroscopy (EDX). The details are laid out in the following paragraphs.

Powder x-ray data was taken with a commercial in-house diffractometer (Rigaku RINT-TTR-III) at $T = 300\,$K, with Cu $K_{\alpha}$ radiation (no monochromator). The software package RIETAN \cite{Izumi2007} was used for the refinement procedure. We present full Rietveld refinement of powder x-ray spectra obtained from crushed single crystal pieces of slightly doped Gd$_2$PdSi$_3$ in Fig. \ref{fig:si_xrd}. The x-ray data was refined with the basic AlB$_2$-type structure, neglecting weak additional reflections due to the formation of a complex, yet still inversion symmetric, crystallographic superstructure related to ordering on the Pd/Si sublattice \cite{Kotsanidis1990, Tang2011, Kurumaji2019}. In contrast, our band structure calculations (section \ref{sec:bstruct}) model the effect of Pd/Si order using a larger Ce$_2$CoSi$_3$ type structure \cite{Kotsanidis1990}. Both AlB$_2$ and Ce$_2$CoSi$_3$ structures correspond to space group $\#191$ (P6/mmm). The even weaker effect of stacking along the $c$-axis \cite{Tang2011, Kurumaji2019} is neglected both in the present refinement of the powder x-ray patterns, and in the band structure calculations. 

The refined lattice parameters show monotonous evolution on the Rh-doped side (Fig. \ref{fig:si_lattice_constants}). The change of lattice parameters is comparatively small in these doping series, when considering absolute units; this is true not only for $a$, but also for $c$ [Fig. \ref{fig:si_lattice_constants}]. We caution that using RIETAN, the patterns can be refined with only marginal changes to the goodness of fit ($S$-value), when adjusting $a$ or $c$ by up to $\pm 0.0015$. This however is not reflected by the statistical errors returned by the program (indicated in Fig. \ref{fig:si_lattice_constants}).

\section[Characterization of crystal structures by EDX]{Characterization of dopant concentration by energy-dispersive x-ray spectroscopy (EDX)}

\begin{figure}[htb]
  \centering
  \includegraphics[width=1.\linewidth]{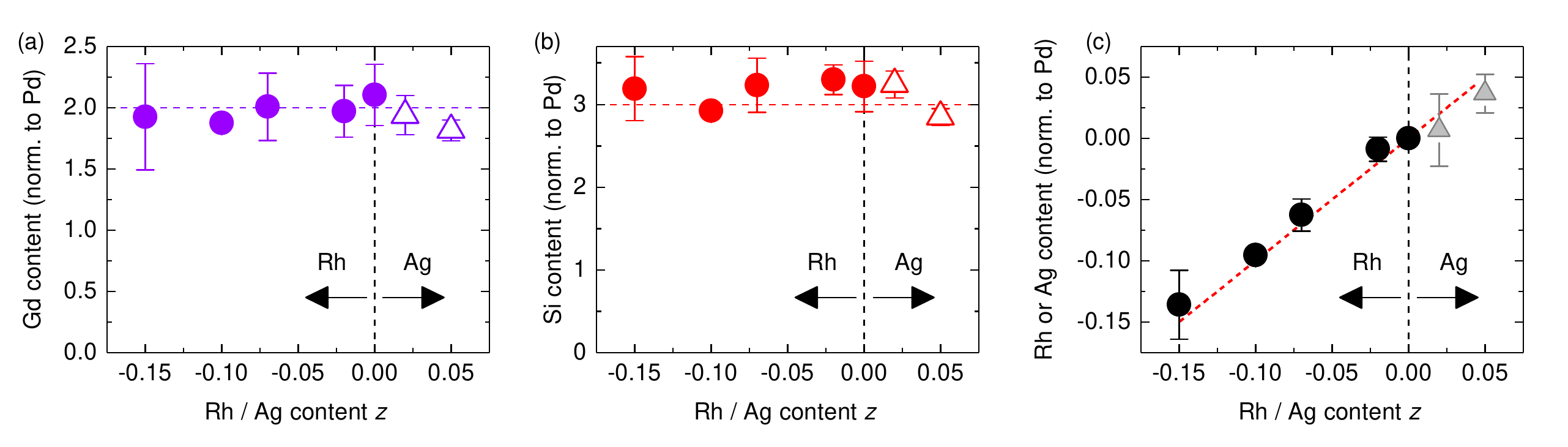}
  \caption{Gadolinium (a), silicon (b), and Rh/Ag (c) atomic content in doped Gd$_2$PdSi$_3$ single crystals. As EDX provides relative atomic fractions, these absolute numbers were obtained by normalizing to the nominal Pd content, e.g. to $y_\text{Pd} = 1-x = 0.85$ for Gd$_2$(Pd$_{1-x}$Rh$_x$)Si$_3$, $x = 0.15$. Black vertical dashed lines indicate $z=0$. Horizontal dashed lines in (a,b) and the tilted line in (c) indicate expected behavior of the atomic constituents (see text).}
\label{fig:si_edx_concentration}
\end{figure}
We used energy-dispersive x-ray spectroscopy (EDX) to study the chemical composition and homogeneity of our single crystals in a scanning-electron microscope equipped with an EDX analyzer (SEM-EDX; JEOL and Bruker). Figure \ref{fig:si_edx_concentration} shows quantitative analysis of the spectra, which was obtained using ESPRIT2 Microanalysis software (Bruker). The error bars in this figure correspond to statistical uncertainty, determined from the standard deviation when measuring the spectra in about $10\,(\pm2)$ different areas of the same sample. We return to the question of the systematic error below. 

Let us focus on Fig. \ref{fig:si_edx_concentration} (c), where the nominal stoichiometry $z$ is equal to $x$ on the Ag-doped side, and to $-x$ on the Rh-doped side. We report the observed stoichiometry $z_m$ as a function of $z$. The expected relation ($z_m = z$) is indicated by the red dashed line in the plot. Correspondingly, Fig. \ref{fig:si_edx_concentration}(a,b) shows data of observed silicon and gadolinium content in our samples, when normalized to the nominal palladium concentration. We discuss in the following why the ESPRIT2 software may struggle with quantitative analysis on the Ag-doped side, and manually carry out a 'sanity check' for these EDX results.

\begin{figure*}[htb]
  \centering
  \includegraphics[trim=2.5cm 1.cm 2.8cm 0.5cm, width=1.0\linewidth]{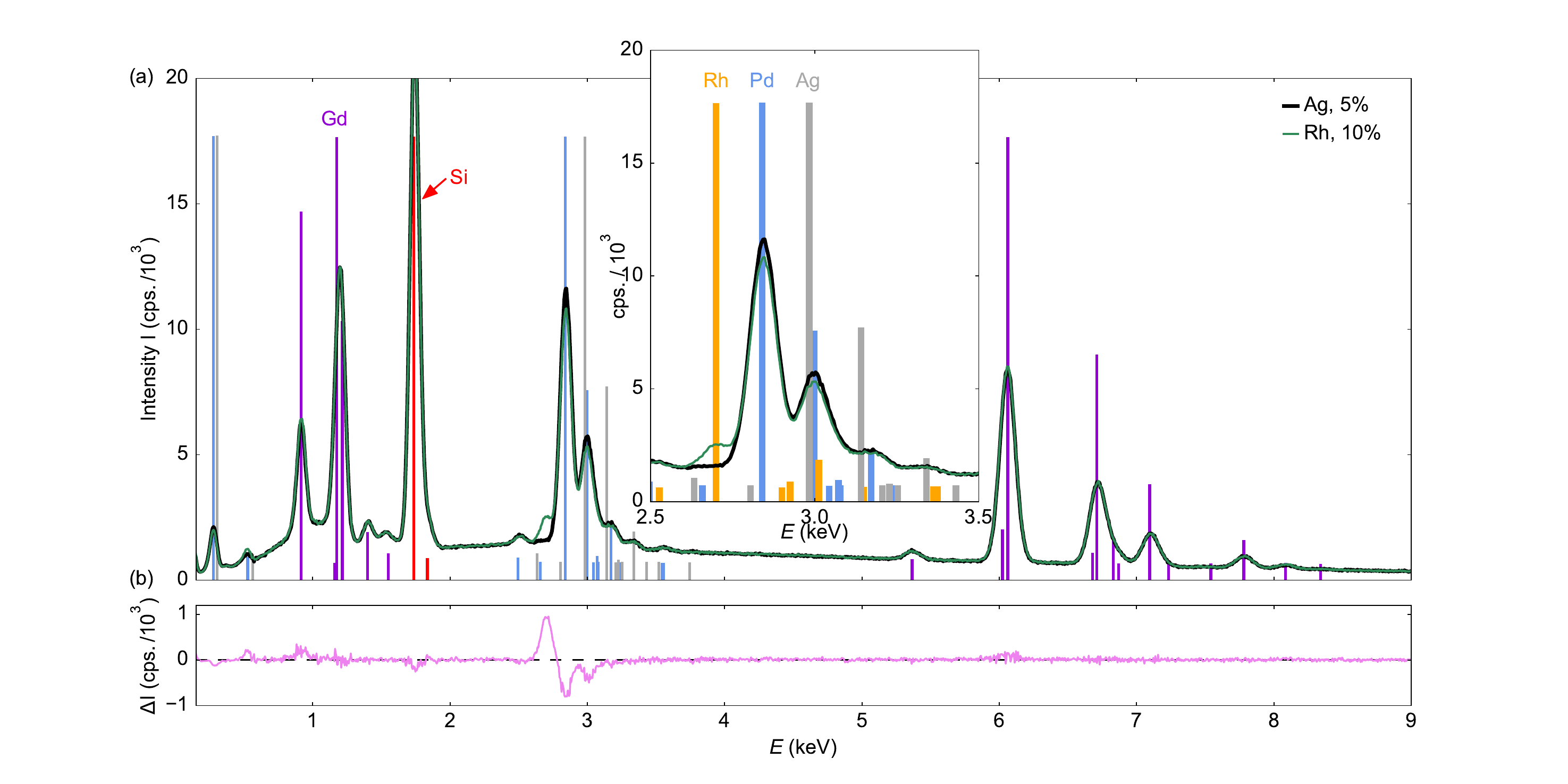}
  \caption{(a) Energy-dispersive x-ray spectroscopy for polished single crystals of Gd$_2$(Pd$_{1-x}$Ag$_x$)Si$_3$ (sample A, black line, $x = 0.05$) and Gd$_2$(Pd$_{1-x}$Rh$_x$)Si$_3$ (sample B, green line, $x = 0.10$, scaled by factor $0.93$ to match). Expected resonances corresponding to Gd, Pd, Si, Ag, and Rh atoms are indicated by violet, light blue, red, grey, and orange bars, respectively. In the inset, expanded view of the regime of the spectrum which is characteristic of the $d$-metals. The Rh-peaks are marked only in the inset, to avoid overcrowding in the main panel. (b) Difference of the two curves in (a), revealing a decrease of the Pd-related signal when comparing $10\%$ Rh-doped and $5\%$ Ag-doped samples. Separation of the Ag-related intensity is difficult due to overlap with a Pd side-peak.}
\label{fig:si_edxpanel}
\end{figure*}

Figure \ref{fig:si_edxpanel} shows representative spectra for two compounds Gd$_2$(Pd$_{1-x}$Ag$_x$)Si$_3$ (sample A, black line, $x = 0.05$) and Gd$_2$(Pd$_{1-x}$Rh$_x$)Si$_3$ (sample B, green line, $x = 0.10$). The expected lines for each element are indicated in the plot. No additional lines were found in the data, indicating that there was no contamination of the samples with another element within the resolution of this technique. Moreover, the inset of Fig. \ref{fig:si_edxpanel}(a) shows an expanded view of the characteristic peaks of the $4d$-metals Pd, Rh, and Ag. From the position of the colored bars it is clear that intensities of Rh and Pd can be easily separated. However, this is less straightforward for the case of Pd and Ag, due to overlap of a secondary Pd peak with the primary Ag peak.

We discuss the origin of the change of error bar size for different data points in Fig. \ref{fig:si_edx_concentration}. Firstly, for Gd$_2$(Pd$_{0.95}$Ag$_{0.05}$)Si$_3$, the contribution of Pd to the peak at $3.0\,$keV is around $90\%$ when making several rough approximations. A characteristic relative (systematic) error of $\pm 1.5\%$ for the detection of Pd thus translates to an error of $\pm 14\%$ in the observed Ag-content. Whereas we assumed that the statistical error dominates on the Rh-doped side, we have combined this systematic error with the aforementioned statistical error on the Ag-doped side. Hence, the error bars in Fig. \ref{fig:si_edx_concentration} are considerably much larger on the Ag-doped side, as compared to the Rh-dope side. We reiterate that the cause of this behavior is line overlap of Pd and Ag. Also note that for the samples with $x_\text{Rh} = 0.15$ and $x_\text{Ag} = 0.02$, only four regions of the crystal were measured in EDX and averaged; thus, the standard error of the statistical distribution of Rh / Ag content is significantly much broader for these samples.

We further check the result of the ESPRIT2 analysis on the Ag-doped side through a 'manual' estimate. In Fig. \ref{fig:si_edxpanel}, data for sample B was scaled by a linear factor of $\times0.93$, so that it matches sample A over a wide interval of energies. The scaling adjusts for slight differences in detector position, beam intensity, and surface scattering (beam loss) between the two measurements. We define the difference $\Delta I = I(\text{sample B})-I(\text{sample A})$ and find significant discrepancies between the two curves [Fig. \ref{fig:si_edxpanel}(b)] only between $2.5$ and $3.2\,$keV, where the signal from the $4d$-metals dominates. Thus, within the resolution of our EDX experiment, the (relative) Gd and Si content of these samples is identical. 

Furthermore, we can obtain information from the absolute magnitude of $\Delta I$. The main Pd peak at $E = 2.84\,$keV is stronger for sample A, as expected due to the smaller dopant concentration. Let us consider the relative change of the Pd peak intensity $\eta_\text{Pd} = \Delta I / \left(I(\text{sample A})-I_\text{bg}\right)$ at $E = 2.84\,$keV, where $I_\text{bg}$ is the smooth (polynomial) background generated by x-ray Bremsstrahlung. The data in Fig. \ref{fig:si_edx_concentration} yields $\eta_\text{Pd} = -0.078$, consistent with $x'_\text{Ag} = 0.035$ for sample A with nominal Ag-concentration of $x_\text{nom., Ag} = 0.05$. This estimate is consistent with the quantitative result returned by the ESPRIT2 software ($x'_\text{Ag} = 0.037$).

\section[Measurement setup]{Experimental setup for high-resolution (thermo-)electric measurements}
\begin{figure*}[htb]
  \centering
  \includegraphics[trim=1.0cm 5cm 3cm 1cm, width=1\linewidth]{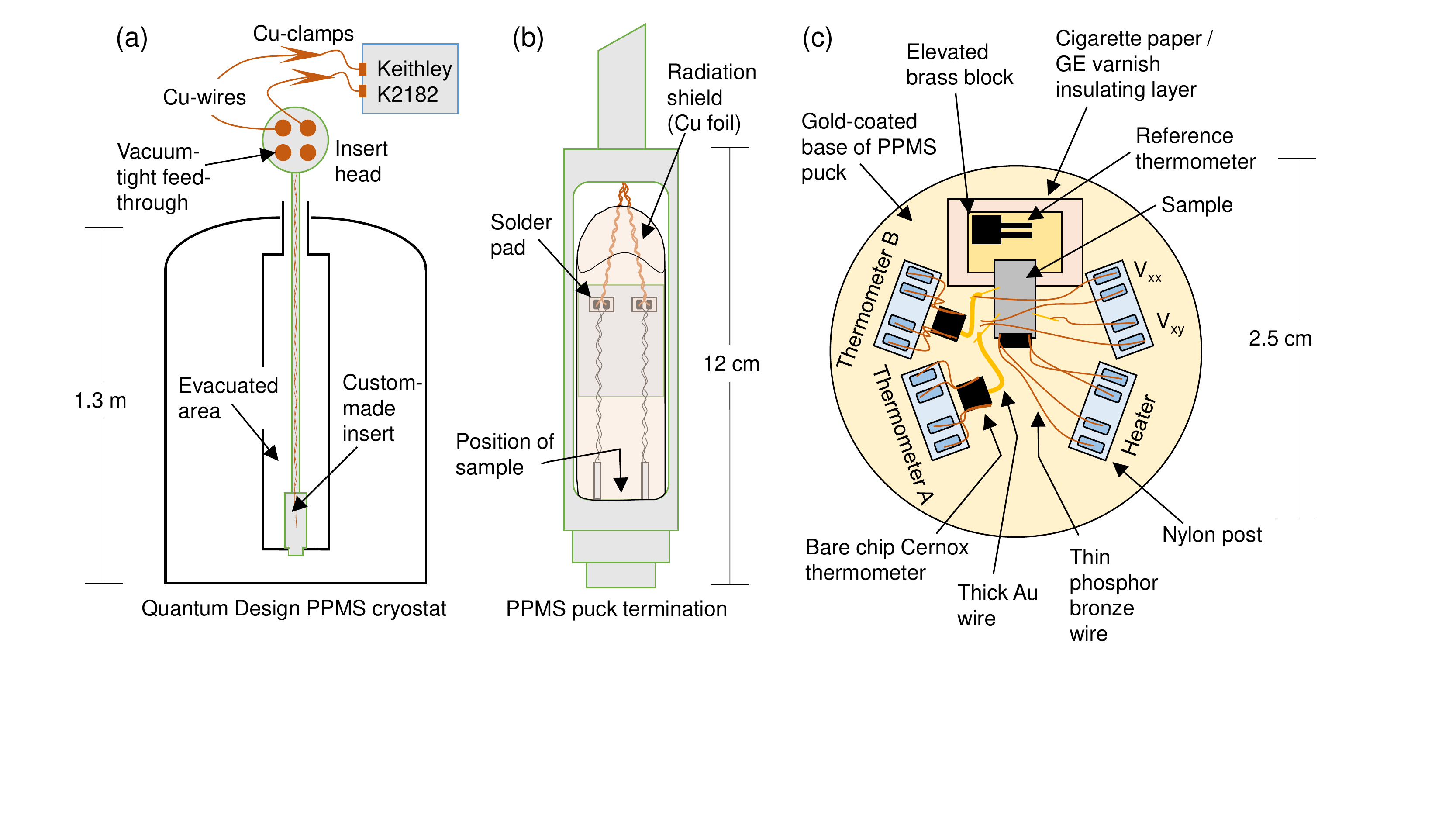}
  \caption{Experimental setup for measurements of Nernst effect and thermopower in Gd$_2$PdSi$_3$. (a) Profile view of the cryostat and custom-made insert for Nernst measurements. Continuous copper wires (red lines), passing through a vacuum-tight wire feedthrough, are clamped directly to the input leads of a Keithley K2182 multimeter. (b) Magnified profile of the low-temperature stage, where twisted pair copper leads (red) arriving from the top are soldered to thinner phosphor bronze wires (black, with insulation). Copper foil is used as a radiation shield. (c) Sample stage, top-view. The sample (grey) is attached to a brass block (yellow) with silver paint, which in turn is heat sunk to the PPMS puck using an insulating junction made of cigarette paper and GE varnish glue. Bare chip thermometers (black squares) are suspended from nylon posts with thin phosphor bronze wires (brown). They are connected to the sample via a combination of thin and thick Au wires (yellow lines), ensuring both good thermal contact and small contact area. The phosphor bronze wires (red) for measuring the sample voltages $V_{xx}$ and $V_{xy}$ are attached to the Au wires emanating from the sample. On the upper side of the nylon posts, we created small contact pads with silver epoxy (dark blue).}
\label{fig:si_nernst_setup}
\end{figure*}
A sketch of the experimental configuration for thermoelectric transport is shown in Fig. \ref{fig:si_nernst_setup}. We used a commercial Quantum Design Physical Properties Measurement System (PPMS) cryostat and developed a custom low-temperature insert suitable for thermoelectric experiments at $T > 7\,$K. The sample was attached with silver pain to a brass block, which was in turn connected to the heat bath through an insulating contact made of GE varnish glue and cigarette paper. This design, including an insulating connection with a large area cross-section and good thermal conductance, allows us to generate sizable temperature gradients without raising the sample's temperature too much. 

Small chip thermometers (Cernox CX1030 bare chip, Lakeshore cyrotronics) and phosphor bronze wires of diameter $40\,\mathrm{\mu m}$ were connected to the sample. These were heat sunk carefully to the probe and extended to the top of the insert. We avoided Pb/Sn solder connections at room temperature, as these are known to generate large, spurious voltage backgrounds. Instead, the connectors of Keithley K2182 voltmeters were clamped directly to the copper wires emerging from the probe. Through thermal shielding in the area of the room-temperature copper clamp connection, care was taken that the temperature at the Cu-Cu joint was relatively stable throughout the experiment. 

In DC voltage measurements of thermopower and Nernst effect, spurious background voltages in the circuit can never be completely eliminated. A solution employed in this study is the delta-mode technique, i.e. applying heat to the sample in pulses of $40-200\,$s duration. The heat pulse is followed by a background measurement of the same time duration, and the background voltage is then subtracted from the signal. In contrast to AC techniques, this approach allows for full thermal equilibration of the sample while also providing a means for a simultaneous background measurement. In summary, the technique presented here is capable of producing signal-to-noise ratios of $2-5\,$nV peak-to-peak, when averaging data taken over $10$ seconds or less, and with minimal systematic errors from spurious voltages in the circuit.

We attached four phosphor-bronze leads to the thermometers for a four-point measurement of the resistance. Lakeshore Cryotronics LS372 temperature controllers were used for the experiments, and the power supplied through the leads was carefully controlled to avoid spurious heating. The thermometers were calibrated in-situ using a reference chip (CX1010, SD, Lakeshore Cryotronics) to overcome tiny resistance shifts which typically occur for these bare-chip sensor after thermal cycling to and from room temperature.

A measurement of the temperature gradient using thermometers is to be preferred to thermocouple-based experiments. This is especially true at low $T<50\,$K, where significant heating of the sample is common in thermoelectric studies. The thermocouple-approach does not allow to easily detect the increase of absolute sample temperature, which can be as large as several Kelvin in our experiment - depending on the heater power ($P\leq 1\,$mW), thermal linking, and heat bath temperature.

Identical sample geometry and wiring were used for both thermoelectric and electric transport measurements. In the case of longitudinal resistivity and Hall effect, we prefer Stanford Research SR830 lockin amplifiers, together with SR560 preamplifiers. The applied AC current (current density) was $5\,$mA (about $3.3\cdot 10^4\,\mathrm{A/m}^2$), with excitation frequency chosen in the range $10-20\,$Hz.

\begin{figure*}[htb]
  \centering
  \includegraphics[width=0.6\linewidth]{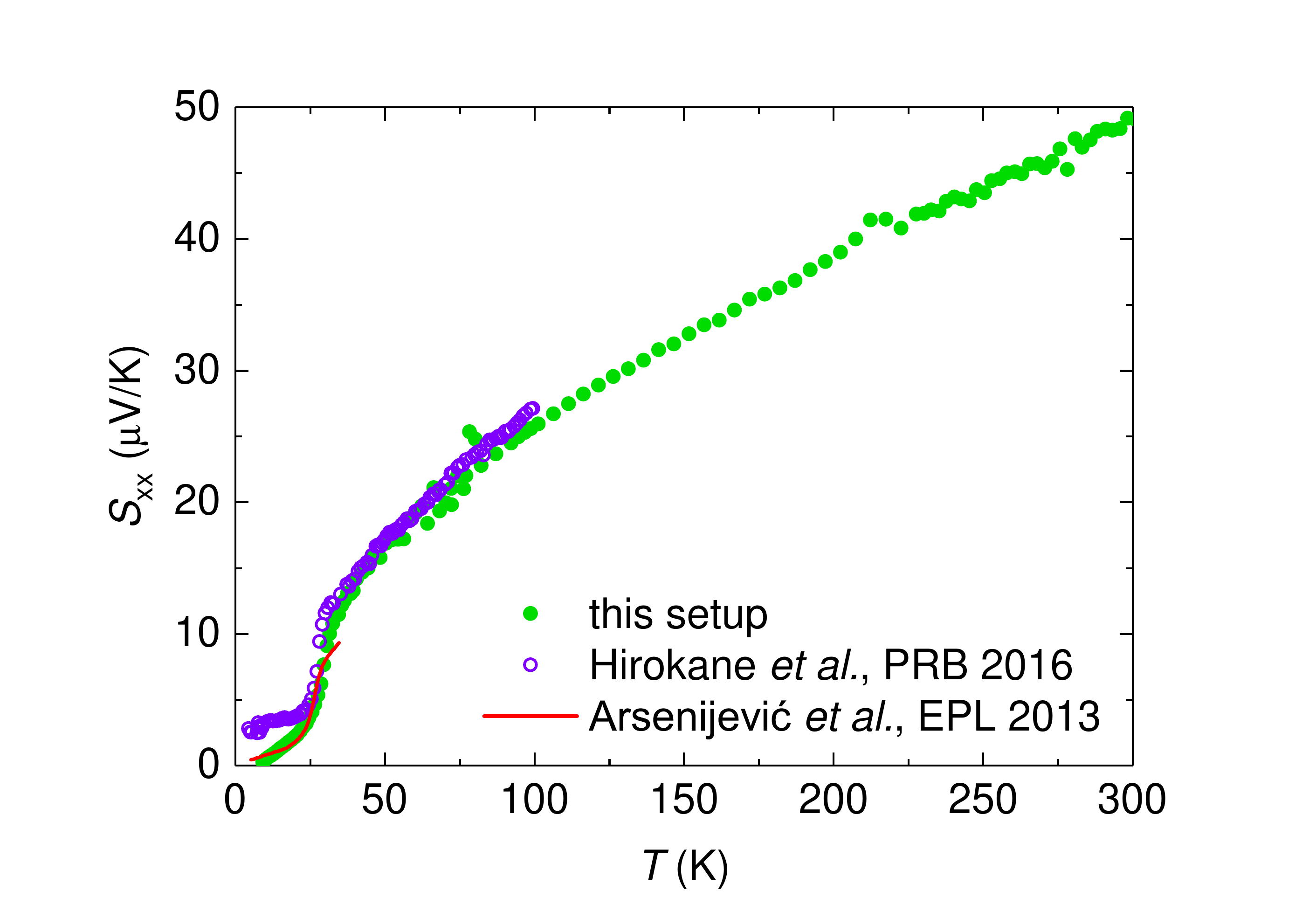}
  \caption{Benchmark test of thermoelectric setup using B20 manganese silicide (MnSi). Temperature dependence of the thermopower was measured in zero magnetic field for $\mathbf{J}_Q\,//\,\,\left[100\right]$. We compare our results with data from the literature. See text for discussion.}
\label{fig:si_sxx_calib}
\end{figure*}
We carried out a benchmark test of our thermoelectric setup using a standard compound, MnSi. Figure \ref{fig:si_sxx_calib} compares thermopower data for single-crystalline MnSi measured using the present setup (green) side by side with data from the literature. The literature data was manually extracted from the respective references:
\begin{itemize}
\item In Hirokane \etal{}, a setup comparable to ours was used, with Cernox thermometers for the measurement of the temperature gradient \cite{Hirokane2016}. 
\item Arsenijevi{\'c} \etal{} measured $S_{xx}$ in a pressure cell, detecting the temperature gradient using thermocouples \cite{Arsenijevic2013}. 
\item Data from the seminal experiment of Ref. \cite{Cheng2010} is not included here, as the disagreement of the absolute value with Fig. \ref{fig:si_sxx_calib} is rather significant. The authors of Ref. \cite{Cheng2010} note that in their own cross-check using an ambient pressure setup, $S_{xx}$ was significantly much larger (several ten percent), but the ambient pressure data is not shown in that manuscript.
\end{itemize}

\section{Band structure calculation}
\label{sec:bstruct}
We perform spin density functional theory (SDFT) calculations using the WIEN2k code \cite{Blaha2020}, where we assume collinear ferromagnetic order and neglect spin-orbit coupling. Since the full crystal unit cell of Gd$_2$PdSi$_3$ is too large due to the complex $c$-axis stacking of different types of Pd-Si layers \cite{Tang2011}, we use the simpler Ce$_2$CoSi$_3$-type crystal structure, which includes four Gd atoms in the unit cell. This simplified model has space group symmetry P6/mmm akin to the full crystal structure, and thus maintains space inversion symmetry. In the calculations of Ag/Rh-doped systems, we stack two Ce$_2$CoSi$_3$-type unit cells along the $c$-axis and replace one Pd atom by Ag/Rh.

The lattice parameters for both pure and doped samples are set to be $a = 8.15676$~\AA\, and $c = 4.09715$~\AA\, (changes of lattice due to chemical substitution are neglected). The internal coordinates of Si atoms are determined by structural optimization. We employ the exchange-correlation functional proposed by Perdew \etal{} \cite{Perdew1996} and set $N_{\bm k}=10^3$ as the number of $\bm k$-points for the self-consistent calculations. Muffin-tin radii ($R_{\rm MT}$) of $2.5$, $2.5$, and $1.85$ Bohr radii for Gd, Pd/Ag/Rh, and Si are used, respectively. The maximum modulus for the reciprocal vectors $(k_{\rm max})$ is chosen such that $R_{\rm MT} \cdot k_{\rm max}=8.0$. The data for the partial density of states, shown in Fig. 4(a) in the main text, are obtained by projection onto the atomic orbitals. $N_{\bm k}=20^3$ as the number of $\bm k$-points and $\Delta=10^{-3}$ Ry as the smearing width were used in the density of states calculations. 

\section[Derivation of the expression for axy]{Derivation of the expression for $\alpha_{xy}$}
The Cartesian frame defined in the main text has $\left(-\nabla T\right)\parallelsum\,\pmb{x}$, $\pmb{B}\parallelsum\,\pmb{z}\parallelsum\,\pmb{c}$, and the $y$-axis perpendicular to these two directions. The \textit{measured} thermoelectric response may be written as (summation implied) $E_i = S_{ij}\partial_j T$, so that the thermopower $S_{xx}$ is positive for hole-like carriers \cite{Ziman1979}. We assume isotropic behavior in the basal plane so that
\begin{equation}
\left(\begin{matrix}E_x \\ E_y\end{matrix}\right) =\left(\begin{matrix}S_{xx} & S_{xy} \\S_{yx} & S_{xx}\end{matrix}\right) \cdot \left(\begin{matrix}\partial_x T \\ \partial_y T\end{matrix}\right)
\end{equation}
The convention is to use $\partial_x T <0$, because temperature $T$ is increasing along the $-\pmb{x}$ direction. Then, $\left|\partial_x T\right| = -\partial_x T$. We further assume that the thermal Hall conductivity is small as compared to the longitudinal thermal conductivity, $\kappa_{xy}\ll \kappa_{xx}$, and write $\partial_y T \equiv 0$. Then, thermopower and Nernst effect may be expressed as
\begin{align}
S_{xx}&=-\frac{E_x}{\left|\partial_x T\right|}\\
S_{xy}&=-S_{yx} = \frac{E_y}{\left|\partial_x T\right|}
\end{align}
In the vortex-Nernst convention (see Refs. \cite{WangPRB2006,OnoseEPL2009} for discussion), the sign of the Nernst coefficient is chosen to be $\text{sgn}\left(e_N\right) = \text{sgn}\left(S_{xy}\right)$, i.e. contrary to the case of the Hall resistivity, the \textit{sign of the Nernst signal} is the sign of the $xy$ component of the respective tensor. Recall that for hole-like Hall effect, $\sigma_{xy}>0$ but $\rho_{yx}>0$ due to the matrix inversion $\bar{\rho} = \bar{\sigma}^{-1}$.

We further derive the relationship between the thermoelectric conductivity tensor defined through $\pmb{J} = \bar{\alpha}\cdot\left(-\nabla T\right)$ and the Nernst signal $S_{xy}$. Under open circuit conditions, the net current in the sample must vanish. Therefore, we expect cancellation of (a) currents driven by the temperature gradient and (b) currents driven by electric fields arising from charge imbalance within the sample:
\begin{equation}
0 = \pmb{J}+\pmb{J}' = \left(\begin{matrix}\alpha_{xx}\left(-\partial_x T\right) + \alpha_{xy}\left(-\partial_y T\right)\\ \alpha_{yx}\left(-\partial_x T\right) + \alpha_{xx}\left(-\partial_y T\right)\end{matrix}\right)+\left(\begin{matrix}\sigma_{xx}E_x+\sigma_{xy}E_y\\ \sigma_{yx}E_x+ \sigma_{xx}E_y\end{matrix}\right)
\end{equation}
Again we ignore all terms $\sim\left(-\partial_y T\right)$. Using the above definitions for the $S_{ij}$, we get a system of equations
\begin{align}
\alpha_{xx}-\sigma_{xx}S_{xx}+\sigma_{xy}S_{xy}&=0\\
\alpha_{yx}-\sigma_{yx}S_{xx}+\sigma_{xx}S_{xy}&=0
\end{align}
From the second equation, it follows immediately 
\begin{equation}
\alpha_{xy} = \sigma_{xx}S_{xy}+\sigma_{xy}S_{xx}
\end{equation}
and eliminating $S_{xx}$ one can write the expression
\begin{equation}
S_{xy} = \rho_{xx}\alpha_{xy}-\rho_{yx}\alpha_{xx}
\end{equation}
Crucially, the sign of $e_N = S_{xy}$ is identical to the sign of $\alpha_{xy}$, if the second term $\sim \sigma_{xy}S_{xx}$ is small as in the present case. Note that whereas in the typical case of $S_{xx}\approx\alpha_{xx}/\sigma_{xx}$, the Mott relation (Eq. (1) of the main text) yields $S_{xx}\propto \left.\partial \ln\left(\sigma_{xx}\right)/\partial\varepsilon\right|_\zeta$, an analogous relation between $S_{xy}$ and $\sigma_{xy}$ cannot be defined even in the most simplified cases.

\begin{figure*}[!htb]
  \centering
  \includegraphics[width=0.95\linewidth]{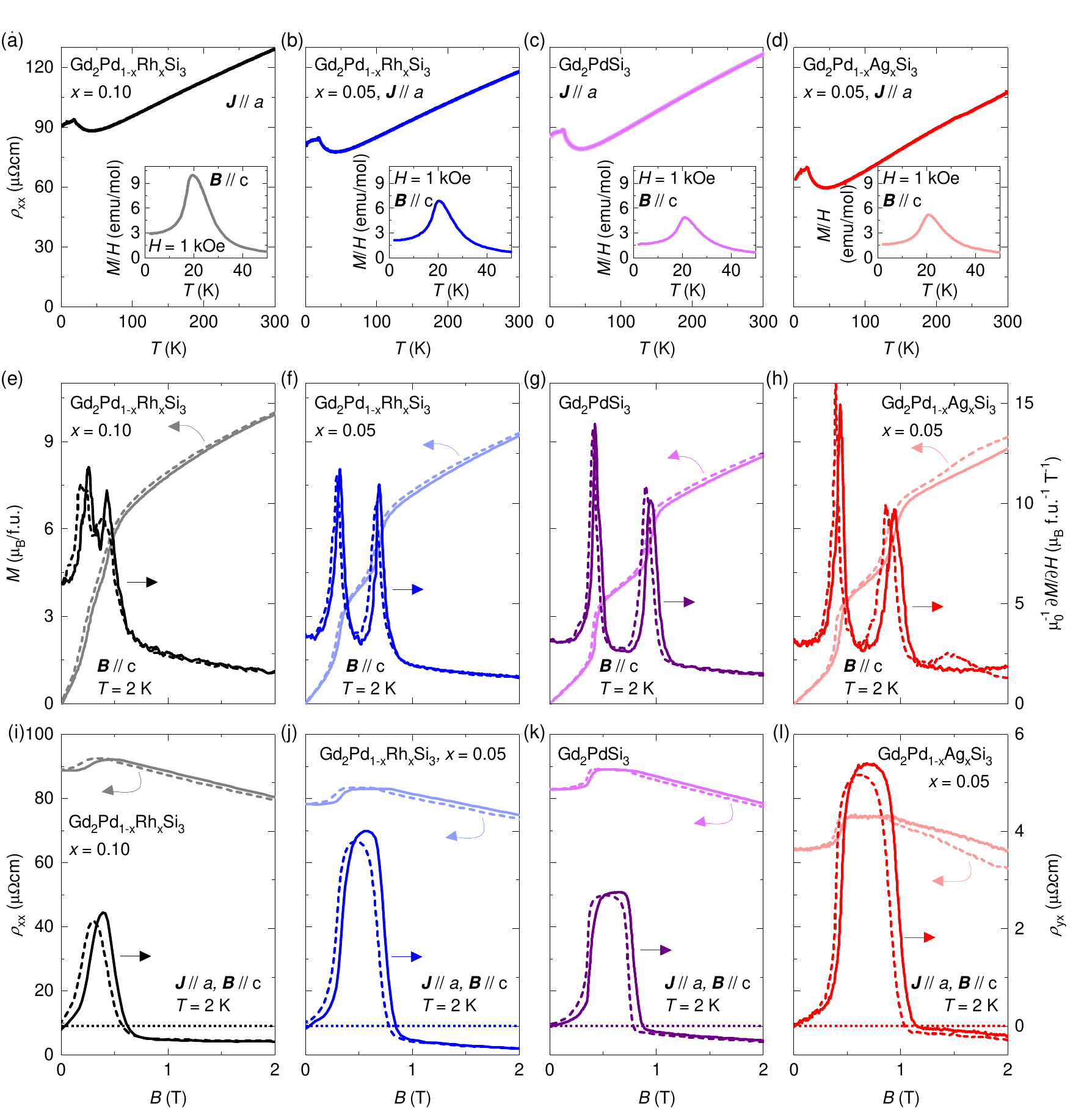}
  \caption{Magnetic and transport properties of electron- and hole-doped Gd$_2$PdSi$_3$. (a-d) Temperature dependence of resistivity $\rho_{xx}$ in zero field shows moderate RRR$\sim 1.5$. The magnetic ordering transition at $T_N$ is observed in all samples (inset). (e-h) Isothermal magnetization $M$ and DC susceptibility $\chi_\text{DC} = \mu_0^{-1} \partial M/\partial H$ at $T = 2\,$K. Sharp first-order transitions delineate the boundaries of the skyrmion lattice state. (i-l) Magnetoresistance $\rho_{xx}(B)$ and Hall resistivity $\rho_{yx}(B)$ at the lowest temperature. A sharp Hall anomaly in the skyrmion lattice phase is present in all samples. The dotted lines mark the zero value of the Hall resistivity. Demagnetization correction was applied to all data.}
\label{fig:si_doping}
\end{figure*}

\section[Comparison of magnitude of anomalous and topological axy in various materials]{Comparison of magnitude of anomalous and topological $\alpha_{xy}$ in various materials}

\begin{figure*}[!htb]
  \centering
  \includegraphics[width=0.95\linewidth]{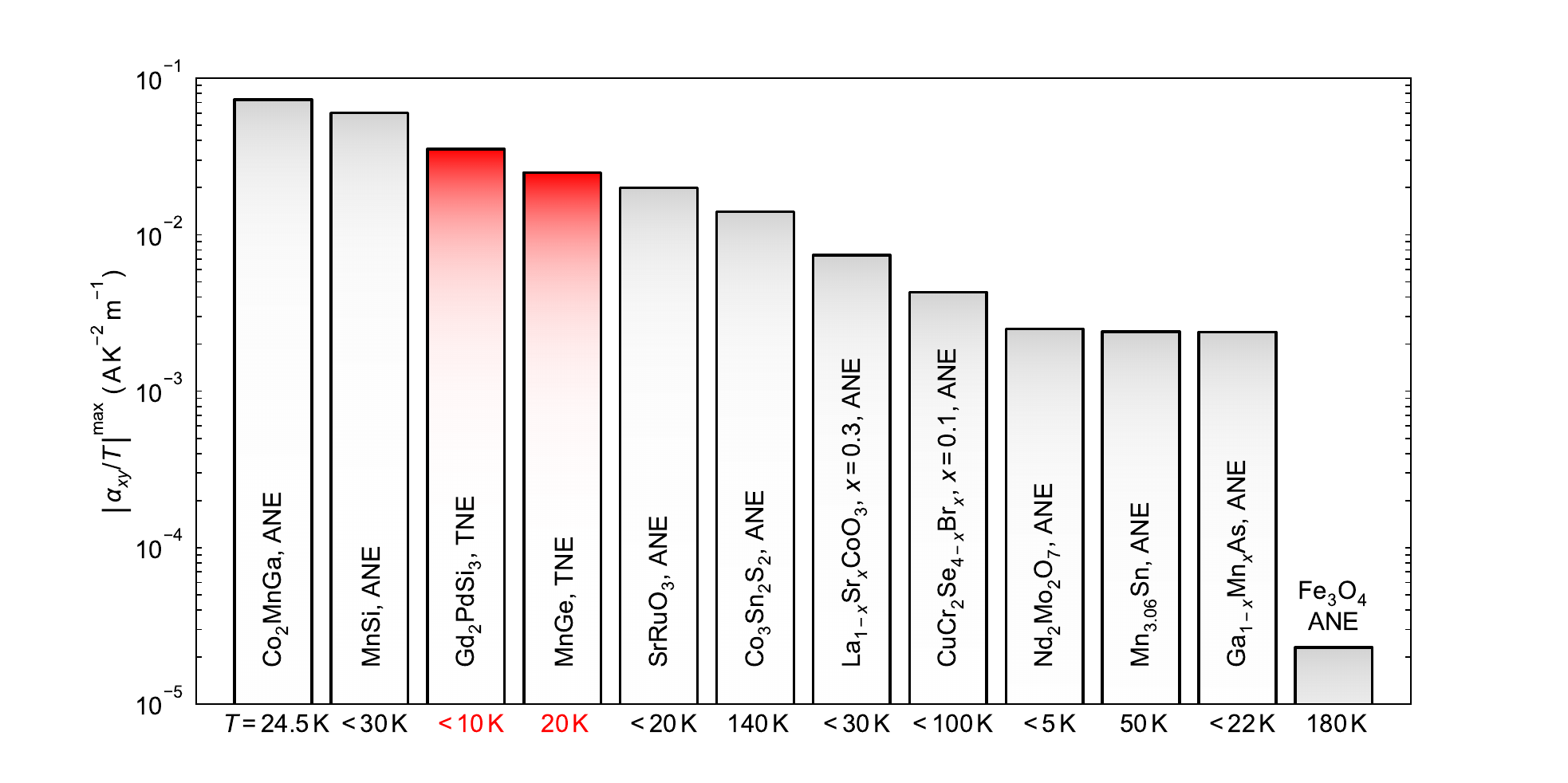}
  \caption{Comparison of magnitude of Nernst conductivity $\alpha_{xy}/T$ (entropy factor removed) for various materials with anomalous (ANE, shaded in grey, proportional to the net magnetization $M$) or topological Nernst effect (TNE, shaded in red). The figure expands the comparison in Fig. 2 (main text) to a larger number of relevant materials. Note the logarithmic $y$-scale as compared to the linear scale in Fig. 2. Temperatures listed below the respective bars refer to the temperature (temperature regime) where the largest value of $\left|\alpha_{xy}/T\right|$ was observed in each case.}
\label{fig:si_materials}
\end{figure*}
Figure \ref{fig:si_materials} shows an extended comparison of $\alpha_{xy}/T$, the transverse thermoelectric conductivity, for compounds where anomalous Nernst effect (ANE) or topological Nernst effect (TNE, in a generalized definition: of magnetic origin, yet not proportional to the magnetization) has been observed in the literature. Data were taken from: Refs. \cite{Sakai2018,Guin2019} for Co$_2$MnGa, Ref. \cite{Hirokane2016} for ANE in MnSi, present work for TNE in Gd$_2$PdSi$_3$, Ref. \cite{Miyasato2007} for SrRuO$_{3}$ and La$_{1-x}$Sr$_x$CoO$_3$, $x= 0.3$ (other $x$ have smaller $\alpha_{xy}/T$), Ref. \cite{Guin2019b} for Co$_3$Sn$_2$S$_2$ (unpublished Ref. \cite{Yang2018} reports enhanced values), Ref. \cite{Lee2004} for CuCr$_2$Se$_{4-x}$Br$_x$, $x = 0.1$ (other $x$ have lower $\alpha_{xy}/T$), Ref. \cite{Hanasaki2008} for Nd$_2$Mo$_2$O$_7$, Ref. \cite{Ikhlas2017} for Mn$_{3.06}$Sn (Mn$_{3.09}$Sn \cite{Ikhlas2017} and 'stoichiometric' Mn$_3$Sn \cite{Li2017} were reported to have slightly smaller values), Ref. \cite{Pu2008} for Ga$_{1-x}$Mn$_x$As ($x=0.07^{*}$ was found to have the largest $\alpha_{xy}/T$), and Ref. \cite{Ramos2014} for Fe$_3$O$_4$.

In the literature comparison above, all ANE signals besides the one observed for MnSi \cite{Hirokane2016} are spontaneous (zero-field) signals. Results obtained from numerical simulations only (e.g. Ref. \cite{Weischenberg2013} for bcc Fe, hcp Co, fcc Ni, and alloys FePd and FePt) were not included.

\section[Resilience of skyrmion lattice phase in doped Gd2PdSi3]{Resilience of skyrmion lattice phase in doped Gd$_2$PdSi$_3$}

Figure \ref{fig:si_doping} presents an overview of the low-temperature magnetic and transport properties of Gd$_2$(Pd$_{1-x}M_x$)Si$_3$ with $M =\,$Rh, Ag at the doping levels discussed in the main text. Overall, we found that the characteristic properties of the transition to long-range order at $T_N \sim 20\,$K remain intact with doping [Fig. \ref{fig:si_doping}(a-d)]. Metamagnetic transitions are smeared out in the most highly doped sample ($x = 0.10$, $M=\,$Rh) but the qualitative features of the $M(B)$ curves are unaffected. Bounded by two sharp peaks in the DC susceptibility $\chi_\text{DC} = \mu_0^{-1}\partial M/\partial H$, the SkL phase was observed in all samples. Here, $\mu_0$ is the permeability of vacuum, and $H$ is the externally applied magnetic field. In transport, the SkL state makes its mark in a gentle enhancement of $\rho_{xx}(B)$ and in the sharp Hall anomaly which, together with the thermoelectric properties, is the focus of the main text. 
%

\section[Resonant elastic x-ray scattering]{Resonant elastic x-ray scattering at Gd-L$_2$ edge}
\begin{figure*}[!htb]
  \centering
  \includegraphics[trim=2.5cm 1.cm 2.8cm 1.cm, width=1.\linewidth]{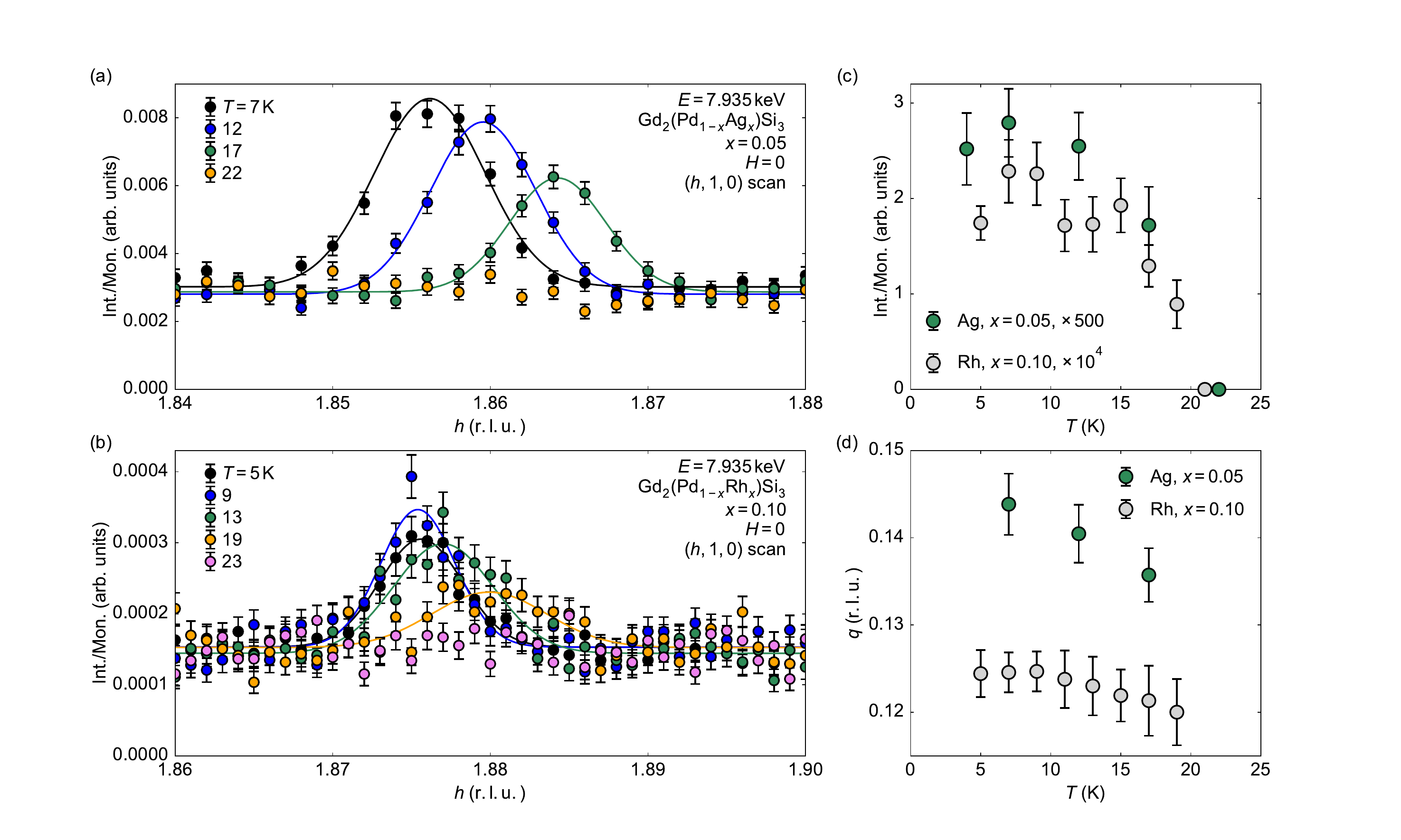}
  \caption{Resonant elastic x-ray scattering (REXS) intensities normalized by monitor counts. (a,b) $(h, 1, 0)$ line cuts, on resonance energy for the Gd-L$_2$ absorption edge, at various temperatures for (a) Gd$_2$(Pd$_{1-x}$Ag$_x$)Si$_3$ ($x = 0.05$) and (b) Gd$_2$(Pd$_{1-x}$Rh$_x$)Si$_3$ ($x = 0.10$). Solid lines correspond to Gaussian fits, with constant background. Error bars correspond to statistical uncertainties. (c) Extracted Gaussian peak amplitude (resonant x-ray scattering intensity) as a function of temperature. (d) Extracted Gaussian center position, corresponding to the magnetic wavenumber $q$ [measured as distance from $(2,1,0)$ Bragg reflection]. Error bars in (c,d) correspond to half-width-at-half-maximum of the Gaussian fit. Statistical error of peak position is even smaller.}
\label{fig:si_rexs_tdep}
\end{figure*}

\begin{figure}[!htb]
  \centering
  \includegraphics[width=0.5\linewidth]{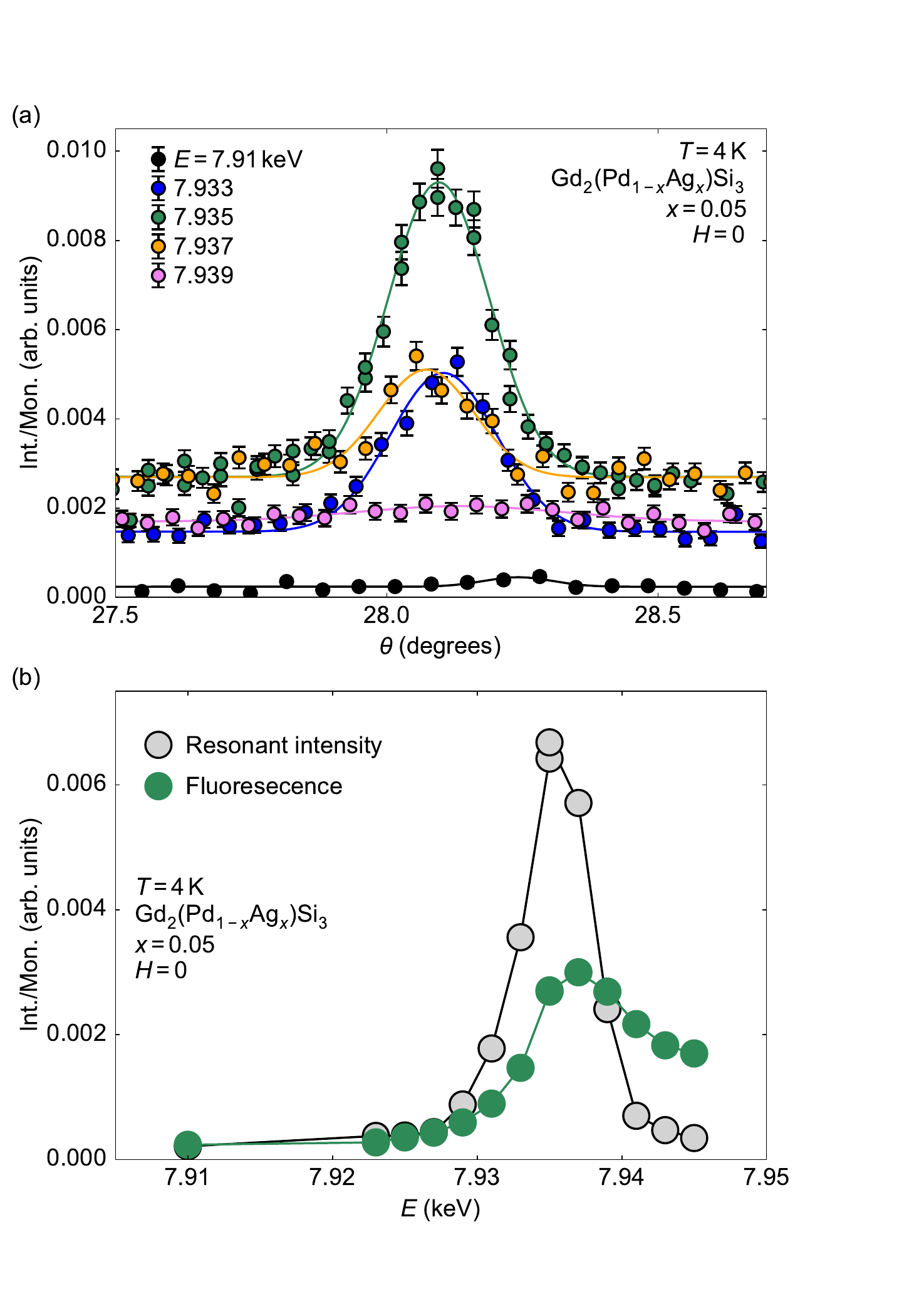}
  \caption{(a) Intensity at the detector divided by monitor counts for $\theta$-scans around $(2-q, 1, 0)$ magnetic peak position at various energies in the vicinity of the resonance energy, for Gd$_2$(Pd$_{1-x}$Ag$_x$)Si$_3$ ($x = 0.05$) at $T = 4\,$K and in zero magnetic field. Solid lines indicate Gaussian fits to the data with constant backgrounds. (b) Extracted Gaussian peak amplitude (resonant scattering intensity) and constant offset (fluorescence signal) as a function of energy. Sharp enhancement of the former is observed around $E_\text{res} = 7.935\,$keV.}
\label{fig:si_rexs_edep}
\end{figure}
Resonant x-ray scattering experiments at the Gd-L$_2$ edge were performed at BL-3A of Photon Factory (KEK, Tsukuba, Japan). Polished single crystals of dimensions $0.8\times 0.85\times 0.5\,$mm$^3$ were mounted in a helium cryostat. We first identified the position of the magnetic peak in momentum space by performing line scans along several high-symmetry directions with energy tuned to the resonant condition $E_\text{res} = 7.935\,$keV. $E_\text{res}$ was known from previous work on the stoichiometric compound\cite{FrontzekPhD, Kurumaji2019}. Spurious backgrounds were identified by repeating the line scans at a slightly detuned energy. 

Subsequently we studied the temperature dependence of the magnetic reflection, centering the sample for each new temperature step using nuclear Bragg peaks. Figure \ref{fig:si_rexs_tdep} shows the results of this experiment, demonstrating vanishing resonant magnetic scattering in the paramagnetic state ($T>20\,$K). Moreover, the $q$-position of the resonant reflection does not change dramatically as a function of dopant concentration [see main text Fig. 3(b)]. The absolute magnitudes of the scattering intensity for Rh- and Ag-doped samples are dissimilar due to differences in the monitor setting. 

The $T = 4\,$K data for Gd$_2$(Pd$_{1-x}$Ag$_x$)Si$_3$ shown in the main text [Fig. 3(b)] is not reproduced in Fig. \ref{fig:si_rexs_tdep}(a). The reason is that the position of the beam on the sample surface was slightly changed between two experimental runs. Therefore, the curve reported for the Ag-doped sample in the main text [Fig. 3(b)] was recorded under slightly different conditions as compared to Fig. \ref{fig:si_rexs_tdep}(a). While the $q$-values extracted from both measurements are in excellent agreement, the intensities are slightly different. 

For Gd$_2$(Pd$_{1-x}$Ag$_x$)Si$_3$, the energy dependence of the magnetic reflection was studied in detail (Fig. \ref{fig:si_rexs_edep}). Similar to the case of Gd$_2$PdSi$_3$ \cite{Kurumaji2019}, strong enhancement of the magnetic intensity was observed only in a very small window of $E$ corresponding to the Gd-L$_2$ absorption edge. Both fluorescence and magnetic signal show the expected behavior [Fig. \ref{fig:si_rexs_edep} (b)]. For Gd$_2$(Pd$_{1-x}$Rh$_x$)Si$_3$, the presence of magnetic scattering was confirmed by suppression of intensity with temperature (Fig. \ref{fig:si_rexs_tdep}) as well as a $(h,1,0)$ line scan at $E' = 7.931\,$keV (not shown), which demonstrates suppression of the magnetic intensity when going off-resonance.

\section[Estimate for energy broadening in Gd2PdSi3]{Estimate for energy broadening in Gd$_2$PdSi$_3$}
Gd$_2$PdSi$_3$ is a system with moderate residual resistivity, as evidenced by the $\rho(T)$ curves in Fig. \ref{fig:si_doping}(a-d). We estimate the uncertainty in the chemical potential $\zeta$ from the scattering time $\tau$, and compare it to the change in Fermi energy $\varepsilon_F = \zeta(T=0)$ expected from introduction of $5-10\,\%$ Rh or Ag dopants on the Pd site. 

In a single-band system, the Hall angle $\tan\theta_H = \sigma_{xy}/\sigma_{xx} = \mu B$ is described by the carrier mobility $\mu = e\tau/m^*$ with electron charge $e>0$, scattering time $\tau$, and carrier effective mass $m^*$. In a multi-band system, the Hall conductivity is typically reduced due to cancellation of Hall currents of opposite sign, and $\tan\theta_H$ can be used to provide a lower bound for $\tau$. For Gd$_2$PdSi$_3$, we estimate $\mu \ge 10\,\mathrm{cm^2\,V^{-1}\,s^{-1}}$ (section \ref{sec:mat_param}) and using $m^*=m_e$ with electron mass $m_e$, we get $\tau >5.5\,$fs. Then, energy-time uncertainty prescribes $\delta \zeta \le \hbar/\tau \sim 100\,$meV, which (we note) is larger than the broadening $1\,\mathrm{mRy} = 13.6\,$meV applied in the band structure calculations of Fig. 4 (main text) and section \ref{sec:bstruct}.

Meanwhile, the $\gamma$-coefficient of the specific heat in the non-magnetic analogue Y$_2$PdSi$_3$ is \cite{Mallik1996}
\begin{equation}
\gamma = \frac{\pi^2}{3}\,k_B^2\,\mathcal{N}_F = 9.0\,\mathrm{mJ\,K^{-2}\,mol^{-1}}
\end{equation}
where $k_B$ and $\mathcal{N}_F = \left(\partial n/\partial \varepsilon\right)_{\zeta}$ are the Boltzmann constant and the density of states at the chemical potential, respectively. From these values, we deduce that $\varepsilon_F = \zeta(T=0)$ shifts by $\sim 13\,$meV when changing the band filling by $\pm 0.05$ electrons per formula unit. 

In summary, the estimate motivates the linear fit of $\sigma_{xy}^\mathrm{max}$ as a function of $x$ in the range $M=\,$Rh, $x = 0.15$ and $M=\,$Ag, $x = 0.05$ [Fig. 3(f)]. The energy broadening $\delta\zeta$ due to impurity scattering is comparable to the change in $\varepsilon_F$ from the additional carriers introduced in the doping experiment. 

\section[Previous tests of the Mott relation for the anomalous Nernst effect]{Previous tests of the Mott relation for the anomalous Nernst effect}
While the Mott relation for the TNE has never been established previously, the case of the anomalous Nernst current has, to our knowledge, been examined in three material systems:

\subsection{Spinel ferromagnet}

Seminal work on the ferromagnetic spinel CuCr$_2$Se$_{4-x}$Br$_x$ reported Hall conductivity $\sigma_{xy}$ and Nernst conductivity $\alpha_{xy}$ as a function of dopant concentration $x$ \cite{Lee2004,Lee2004b}. Note that both carrier concentration $n_h$ (carriers are hole-like) and scattering time $\tau$ vary significantly with $x$. In Ref. \cite{Lee2004}, the relation $\left|\rho_{yx}^A\right|/n_h = A\rho_{xx}^\alpha$ with anomalous Hall resistivity $\rho_{yx}^A$ and resistivity $\rho_{xx}$ was reported. The parameters were found to be $\alpha = 1.95\pm 0.08$ and $A = 2.24\cdot 10^{-25}\,$ (SI) \cite{Lee2004}. 

Meanwhile, Ref. \cite{Lee2004b} used the expression $\alpha_{xy}^A/T =\mathcal{A}\frac{e' k_B^2}{\hbar}\mathcal{N}_F$, where $\mathcal{N}_F = \left(\partial n_h/\partial \varepsilon\right)_\zeta$ is the density of states (DOS) at the chemical potential / Fermi energy, $e'$ (here chosen to be $<0$) is the charge of the electron, $k_B$ is the Boltzmann constant, and $\mathcal{A} = 33.8\,\mathrm{\AA}^2$ denotes a microscopic area related to the Berry curvature. In their analysis, the authors of Ref. \cite{Lee2004b} set $\mathcal{N}_F = \mathcal{N}_F^{(0)}$ to be the density of states of the free electron gas.

Returning to the explicit form of the Mott relation and inserting the expression $\left|\sigma_{xy}^A\right| \approx \left|\rho_{yx}^A\right|/\rho_{xx}^2 = An_h$ from Ref. \cite{Lee2004},
\begin{equation}
\frac{\alpha_{xy}^A}{T} = \frac{\pi^2}{3}\,\frac{k_B^2}{e'}\,\left(\frac{\partial \sigma_{xy}^A}{\partial \varepsilon}\right)_\zeta = \frac{\pi^2}{3}\,\frac{k_B^2}{e'}\,A\,\mathcal{N}_F^{(0)}(\zeta)
\end{equation}
where we continue using $e'<0$. It follows that $\mathcal{A} = A\,\left(\pi^2/3\right)\,\left(\hbar/e'^2\right)$ within the conventions of Lee \etal{}. The right side of this equation gives a value of $3.03\cdot 10^{-21}\,$m$^2$. This means that there is a significant discrepancy with the value of $\mathcal{A} = 3.38\cdot 10^{-19}\,$m$^2$ extracted in Ref. \cite{Lee2004b} from the $x$-dependence of $\alpha_{xy}^A/T$ alone. We speculate that the discrepancy arises from the assumptions made about $\mathcal{N}_F$, or from significant changes in the spin polarization as a function of $x$, which are neglected in the above analysis. However, the Mott relation for the anomalous Nernst effect was therefore not confirmed for the spinel ferromagnet within the scope of the analysis reported in the literature.

\subsection{Dilute magnetic semiconductor Ga$_{1-x}$Mn$_x$As}
The authors of Ref. \cite{Pu2008} report a beautiful study on a series of Ga$_{1-x}$Mn$_x$As where magnetism is tuned with manganese content. Although they do not explicitly report the variation of the Hall signal with gating or tuning of the Fermi energy by chemical substitution, they employ the Mott relation in conjunction with the assumption of a power law for the electrical Hall resistivity $\rho_{yx}^A = \lambda M_z \rho_{xx}^n$ to derive (Eq. (2) of Ref. \cite{Pu2008})
\begin{equation}
S_{yx}^A = \frac{\rho_{xy}^A}{\rho_{xx}}\left(T\,\frac{\pi^2k_B^2}{3e'}\,\frac{\lambda'}{\lambda}-(n-1)\,S_{xx}\right)
\end{equation}
where $\lambda' = \partial \lambda/\partial \varepsilon$ is the energy derivative of the proportionality factor $\lambda$, $M_z$ is the component of the magnetization parallel to the external field, and again $e'<0$ is the electron charge. This expression is then fitted to the temperature dependent anomalous Nernst signal, and good agreement is obtained, for various doping levels $x$, with exponents $n = 1.95\pm 0.1$ \cite{Pu2008}. The authors did not report their fit results for $\lambda'/\lambda$, and the validity of the Mott relation was not verified \textit{without adjustable parameters}.

\subsection{Dilute magnetic topological insulator Cr$_x$(Sb$_{1-y}$Bi$_y$)$_{2-x}$Te$_{3}$}
The advantage of the material platform studied by Guo \etal{} is that direct variation of the chemical potential could be achieved, over a large range in energies $\varepsilon$, by gating of thin film samples. This allows for a comparison of $\varepsilon$-dependent changes of the anomalous Hall conductivity $\sigma_{xy}^A$ with the anomalous Nernst signal $S_{xy}^A$, both of which were observed simultaneously. Good quantitative agreement between $S_{xy}^\text{A,calc.}$ - calculated from the gating dependence of $\sigma_{xx}$ and $\sigma_{xy}^A$ - and the measured $S_{xy}^A$ was reported by the authors. However, interestingly, a slight discrepancy between these two quantities arises in close vicinity of the charge-neutral point (Dirac point) \cite{Guo2017}.

\section{Estimate for the depinning threshold of the skyrmion lattice}
The motion of skyrmion lattices can be manipulated by thermal gradients \cite{Everschor2012}. According to Kong and Zang, the spatially dependent population statistics of magnons in a thermal gradient cause a net magnon current \cite{Kong2013}. This magnon current drags the skyrmions towards hotter areas. They estimate that for insulators, where the magnon transport is most efficient due to low Gilbert damping, the velocity of the skyrmion lattice $\mathbf{v}_\mathrm{th}$ can be approximated via 
\begin{equation}
 \mathbf{v}_\mathrm{th} = \frac{T}{J} \frac{\lambda^2}{\alpha M_0^2} \frac{k_\mathrm{B}^2}{24\pi\gamma\hbar^2} \left(\nabla T \right)
\end{equation}
which is in excellent agreement with their micromagnetic simulations. Here $M_0$ is the size of the local magnetic moment and $\lambda=2\pi J/D$ is the period of the helical phase, a measure for the diameter of the skyrmions. Moreover, $\alpha\ll1$ is the Gilbert damping - usually small in insulators due to the excitation gap. For the temperature gradient of $\partial T/\partial x \approx 10^3\, \mathrm{K}/\mathrm{m}$ applied in the present experiment, Kong and Zang estimate the velocity of a skyrmion lattice in the insulator Cu$_2$OSeO$_3$ to be around $0.1\, \mathrm{m}/\mathrm{s}$. 

For the metallic compounds MnSi and Gd$_2$PdSi$_3$, these results should not be taken as quantitative estimates. However, they still give an upper bound for $v_\mathrm{th}$, keeping in mind that the real velocity could be much lower.

For MnSi with $\lambda=18\,\mathrm{nm}$, $M_0=1/2$, and $\alpha=0.1$ we obtain $v_\mathrm{th}^\mathrm{MnSi}\approx1\,\mathrm{mm}/\mathrm{s}$ and for Gd$_2$PdSi$_3$ with $\lambda=3\mathrm{nm}$, $M_0=7/2$, and $\alpha=0.1$ follows $v_\mathrm{th}^\mathrm{Gd_2PdSi_3}\approx0.3\,\mathrm{\mu m}/\mathrm{s}$. For the latter system it is furthermore unclear whether the scaling breaks down in the limit of nanometric skyrmions. In any case, it becomes apparent that the velocity of the skyrmion lattice in Gd$_2$PdSi$_3$ will be very low compared to other, noncentrosymmetric systems such as MnSi or Cu$_2$OSeO$_3$. 

The velocity of the skyrmion lattice due to the thermal gradient has to be compared to the depinning velocity $\mathbf{v}_\mathrm{pin}$. For MnSi, this value has been measured experimentally \cite{Schulz2012} as $v_\mathrm{pin}^\mathrm{MnSi}\approx0.1\,\mathrm{mm}/\mathrm{s}$ and is thus comparable to $v_\mathrm{th}^\mathrm{MnSi}$. Unfortunately, $v_\mathrm{pin}^\mathrm{Gd_2PdSi_3}$ is not known at the present time. Following Hoshino and Nagaosa \cite{Hishino2018}, however, the depinning velocity is proportional to $M_0 \lambda$. By scaling $v_\mathrm{pin}^\mathrm{MnSi}$ accordingly, we get $v_\mathrm{pin}^\mathrm{Gd_2PdSi_3}\approx0.1\,\mathrm{mm}/\mathrm{s}$ as a rough order-of-magnitude estimate. Again, we do not consider this approximation to be quantitative, but it yields the qualitative result $v_\mathrm{pin}^\mathrm{Gd_2PdSi_3} \gg v_\mathrm{th}^\mathrm{Gd_2PdSi_3}$. We can therefore be confident that the skyrmion lattice in Gd$_2$PdSi$_3$ remains pinned in our experiment.

\section[High-field axy in Gd2PdSi3]{High-field $\alpha_{xy}$ in Gd$_2$PdSi$_3$}

\begin{figure*}[htb]
  \centering
  \includegraphics[width=0.9\linewidth]{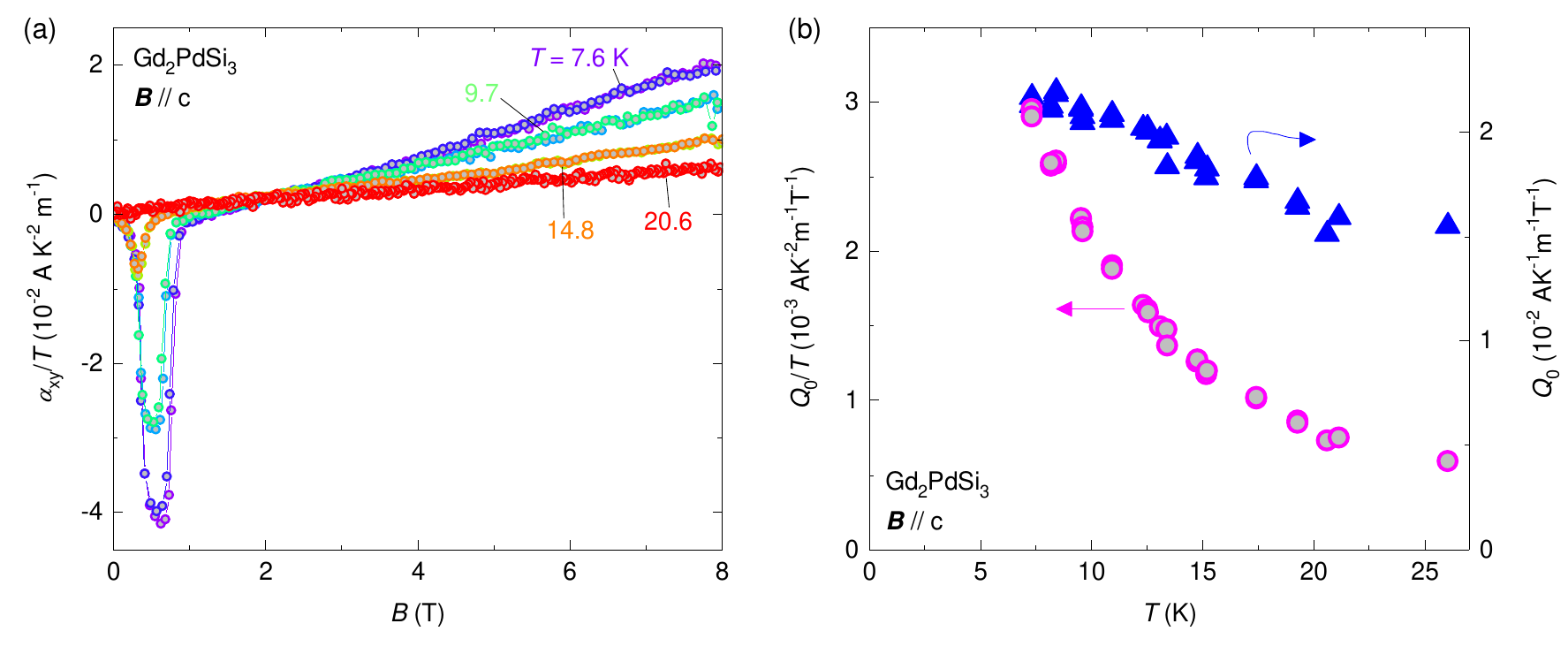}
  \caption{Expanded view of transverse thermoelectric (Peltier) conductivity $\alpha_{xy}/T$ for hexagonal skyrmion host compound Gd$_2$PdSi$_3$. The high-field slope $Q_0$ in panel (b) was obtained through a linear fit to the results in panel (a) in the regime $B>2\,$T. Demagnetization correction was applied to the data.}
\label{fig:si_gd2pdsi3_alphaxy_hfield}
\end{figure*}
Figure \ref{fig:si_gd2pdsi3_alphaxy_hfield} shows a full view of the obtained Nernst conductivity $\alpha_{xy}$ in Gd$_2$PdSi$_3$ including the higher-field regime. We extracted the Nernst coefficient $Q_0$, which is defined as the linear slope of the high-field normal Nernst effect (after demagnetization correction):
\begin{equation}
\alpha_{xy}^{n} = Q_0\cdot B
\end{equation}
We found gentle and rather strong temperature dependence of $Q_0$ and $Q_0/T$, respectively [Fig. \ref{fig:si_gd2pdsi3_alphaxy_hfield}(b)].

\section[Estimates of material parameters]{Estimates of material parameters of SkL Gd$_2$PdSi$_3$}
\label{sec:mat_param}
In the main text, we present estimates for (a) the carrier mean free path and (b) the Hall conductivity in the intrinsic regime. We elaborate on these numbers in the following section.

As for the mean free path $l_\text{mfp} = v_F \cdot \tau$ (Fermi velocity $v_F$, scattering time $\tau$), we adopt an experimentalist's approach, using the available information for the longitudinal resistivity $\rho_{xx}$ and normal Hall coefficient $R_0$ in the high-field regime for stoichiometric Gd$_2$PdSi$_3$. As the Ag-doped samples are somewhat more metallic (Fig. 3, main text), this underestimates the mean-free path in the electron-doped samples by about $50\%$. Note that $\tau$ cannot be determined from the electronic structure calculations. A numerical estimate of $v_F$ remains challenging, because only limited ARPES experiments are available at present for comparison to the numerical 'spaghetti'-type electronic bands \cite{Inosov2009}. Moreover, calculated $v_F$ can severely underestimate the effect of correlations \cite{Taillefer1986}. Meanwhile, integrated properties such as the (partial) density of states (Fig. 4, main text) are expected to be highly robust - both to the onset of magnetic ordering and to the gentle crystallographic superstructure from Pd-Si ordering in single crystals of this material \cite{Tang2011}.

For Gd$_2$PdSi$_3$, $\rho_{xx}(2\,\mathrm{K}) \approx 83\,\mathrm{\mu\Omega cm}$ (Fig. 3, main text) and $R_0 \approx 8\cdot 10^{-10}\,\mathrm{\Omega m / T}$ estimated in the spin-polarized state ($>10\,$T) as part of the present study. We calculate a Hall mobility $\mu_H = R_0/\rho_{xx} = e\tau v_F/(\hbar k_F) = e\,l_\text{mfp}/(\hbar k_F)\approx 9.64\cdot 10^{-4}\,\mathrm{m^2/(Vs)}$. As cancellation of electrons and holes in the Hall effect was neglected, this represents a lower bound for the carrier mobilities and for $l_\text{mfp}$. Further estimating $k_F$ from $R_0$ in a spherical (cylindrical) approximation with two-fold spin degeneracy, we obtain $k_F = 0.61\,\mathrm{\AA}^{-1}$ ($k_F = 0.45\,\mathrm{\AA}^{-1}$), a fraction of $34\%$ ($25\%$) of the reciprocal lattice constant $a^*$. The resulting mean free path is $l_\text{mfp} = 3.9\,$nm ($2.8\,$nm). For comparison, we note that although $l_\text{mfp}$ may reach large values of up to $300\,$nm in ultra-clean samples of the benchmark compound MnSi at the lowest $T$ \cite{Pfleiderer1997}, its value in the regime of the skyrmion lattice is $l_\text{mfp}(28\,\mathrm{K}) \approx 6\,$nm in view of a spin texture with $\lambda_\text{sk} = 18\,$nm.

We would like to emphasize that the magnitude of the topological Hall conductivity observed in Gd$_2$(Pd$_{0.95}$Ag$_{0.05}$)Si$_3$ ($\sim 1000\,\mathrm{\Omega}^{-1}\mathrm{cm}^{-1}$) is comparable to the signals associated with ferromagnets \cite{Miyasato2007,Onoda2008} and coplanar antiferromagnets \cite{Chen2014,Nakatsuji2015} in the intrinsic (momentum-space) limit of intermediate disorder. The characteristic magnitude of $\sigma_{xy}$ in the intrinsic regime is calculated by assuming a single momentum-space monopole per (decoupled) layer. (e.g. Ref. \cite{Ye2018}). In Gd$_2$PdSi$_3$, $\sigma_{xy}^\text{int.} \approx \left(e^2/h\right)\,c^{-1}$, where $c = 4.1\,\mathrm{\AA}$ is the $c$-axis lattice constant. We obtain $\sigma_{xy}^\text{int.}\approx 950\,\mathrm{\Omega}^{-1}\mathrm{cm}^{-1}$, in good agreement with the experimental observation. Note that this estimate by no means implies quantization (or proximity to quantization) for the Hall response: even ferromagnets with extremely complex Fermi surfaces (e.g. single crystals of elemental Fe or thin films of Gd \cite{Miyasato2007}), which have been hypothesized to host a large number of band degeneracy points close to the chemical potential $\zeta$ \cite{Martinez2015}, can show comparable $\sigma_{xy}$.


\begin{thebibliography}{43}
\bibitem [{SI()}]{SI}%
  \BibitemOpen
  \href@noop {} {\enquote {\bibinfo {title} {See {S}upplemental {M}aterial for
  sample characterization, details on thermoelectric measurements, supporting
  resonant x-ray scattering data, and theoretical estimates, which includes
  refs. [31-61]},}\ }\BibitemShut {NoStop}%
\end{thebibliography}
\end{document}